\documentclass[aps,prev,preprint,preprintnumbers,floatfix,doublespace, nofootinbib,13pt]{revtex4}
\usepackage{mediabb}
\usepackage{graphicx,amsmath,amssymb}
\usepackage{color}
\usepackage{mathrsfs}
\usepackage{verbatim}
\usepackage{natbib}

\begin{document}
\vskip.5cm
\title
{Prospect for Study of Randall-Sundrum model from Higgs decay at future colliders}

\author{Hirohisa Kubota$^1$}
\author{Mihoko Nojiri$^{1,2}$} 
\affiliation{
KEK Theory Center$^1$, Tsukuba, Ibaraki 305-0801, Japan \\
The Graduate University for Advanced Studies (Sokendai), \\
Department of Particle and Nuclear Physics, Tsukuba, Ibaraki 305-0801, Japan \\
Kavli IPMU (WPI)$^2$, Tokyo University, Kashiwanoha 5-1-5, Kashiwa, Chiba 277-8583, Japan }

\vglue 0.3truecm

\begin{abstract}
In a class of Randall-Sundrum(RS) models, matter fermions and gauge bosons live in a five dimensional bulk, while the Higgs boson lives in a four dimensional visible brane. 
The Higgs boson can mix with a radion, by a Higgs-radion mixing term.   It is difficult to directly discover the heavy Kaluza-Klein(KK) particles at the Large Hadron Collider(LHC), because the mass of the lightest KK particle is expected to be above 10 TeV to satisfy constraints from flavor changing neutral currents(FCNC).  Instead, the precision measurements of the Higgs sector 
at the  high luminosity LHC(HL-LHC) and the International Linear Collider(ILC) is a promising way to observe the deviation originating from the Higgs-radion mixing and radiative corrections from the KK gauge bosons and matter fermions.  
For some cases, those effects are separately determined, providing valuable information on the model.
We perform an extensive scan of the model parameters to see the expected deviations of the Higgs couplings. 
We also choose several reference points consistent with the current data to show the precision of the fundamental parameter measurements at the future colliders.
We also study the radion-$Z$-$Z$ coupling in the model and discuss the role of the ILC for radion search. 
\end{abstract}.

\maketitle

\section{introduction}
The Standard Model (SM) has been tested in the interaction of various particles very precisely, and so far there is no serious discrepancy between  the SM predictions and experimental data.  In spite of these successes, 
the SM has unnatural features. The Higgs sector of the SM suffers from the gauge hierarchy problem, 
and there are also large hierarchies among the masses of fermions. 
Recently a Higgs boson with a mass of about 125 GeV has been discovered by the ATLAS and CMS experiments~\cite{Aad:2012tfa,Chatrchyan:2012ufa}. The nature of the Higgs boson has not yet been determined precisely. Precision measurements of the Higgs couplings  at the LHC and the proposed ILC might open up a new window to study the effect of physics beyond the SM model (BSM)  in the near future.  

There are several well motivated BSM scenarios for example supersymmetry~\cite{Martin:1997ns}, extra dimension~\cite{Randall:1999ee,ArkaniHamed:1998rs} and composite model~\cite{Contino:2003ve}, aiming to solve the problems in the SM. 
In this paper, we focus on the Randall-Sundrum(RS) model, which is a class of extra dimension models where the matter fermions and gauge fields of the SM live in the bulk~\cite{Goldberger:1999wh,Davoudiasl:1999tf,Pomarol:1999ad,Grossman:1999ra,Gherghetta:2000qt,Chang:1999nh,Davoudiasl:2000wi,Huber:2000fh,Agashe:2003zs,Barger:2011qn,Azatov:2010pf,Goertz:2011hj,Carena:2012fk,Kubota:2012in} while the Higgs boson lives in the four dimensional boundary~\cite{Giudice:2000av,Csaki:1999mp,Csaki:2000zn,Dominici:2002jv,Csaki:2007ns,Chaichian:2001rq,Desai:2013pga,Grzadkowski:2012ng,deSandes:2011zs,Cheung:2003ze}.
The RS model resolves the gauge hierarchy problem by introducing one small warped extra dimension. Due to the cosmological constant of the five dimensional theory, there is a warp factor  $e^{-kr_c \pi}$ in the RS metric, where $k$ is the fifth dimensional curvature scale, and $r_c$ is the fifth dimensional compactification radius. Boundaries of the fifth dimension are two 3-branes which are called the $\lq \lq$visible" and the $\lq \lq$hidden" brane. The mass scale of the visible brane is suppressed by the warp factor while that of the  hidden brane is at the Planck scale.
 
The position of branes may be stabilized by a bulk scalar with boundary interaction terms~\cite{Goldberger:1999uk,Goldberger:1999un}.
In the effective four dimensional theory of the RS-type model, there is a scalar field called $\lq \lq$radion" which corresponds to fluctuations of the size of the extra dimension. The radion obtains a mass due to the stabilization mechanism and can be the lightest new particle in the RS model~\cite{Goldberger:1999un,Csaki:2000zn}.
The  radion and the Higgs boson can mix in the effective four dimensional theory, and thereby 
the couplings of the Higgs boson can deviate  from the SM ones. 

The RS model with bulk fermions relaxes the hierarchies among the SM fermion masses. 
The Yukawa couplings of the effective Lagrangian depend on  the fermion wave functions 
at the visible brane, and  these fermion  profiles are sensitive functions of  the  fermion mass parameters of the five dimensional theory. No hierarchy 
among the fermion masses is needed to reproduce the SM Yukawa couplings~\cite{Grossman:1999ra,Gherghetta:2000qt}. 

In this model, KK modes of the bulk fermions and gauge bosons are predicted in the effective four
dimensional theory~\cite{Goldberger:1999wh,Davoudiasl:1999tf,Pomarol:1999ad,Grossman:1999ra,Gherghetta:2000qt,Chang:1999nh,Davoudiasl:2000wi}. 
The mass of the lightest KK particle has to be above 10 TeV from the kaon mixing constraint~\cite{Csaki:2008zd,Gedalia:2009ws,Blanke:2008zb,Casagrande:2008hr,Bauer:2009cf}, and they are out of reach at any future collider experiments.
Therefore, we focus on the phenomenology of the Higgs sector in the RS model
in this paper. 

In the low energy theory, the radion has some mixing with the Higgs boson.
Moreover, the KK particles of the bulk gauge bosons and matter fermions contribute to the Higgs(radion) couplings to the massless gauge bosons through the  triangle loop diagram .
When the masses of the lightest KK particle is above 10 TeV, large deviations from the SM are not expected from the KK contributions. 
However, the precision of the Higgs coupling measurement is expected to be less than a percent level at the ILC when combined with the HL-LHC data.
By identifying the size of the KK contributions and the Higgs-radion mixing, one may obtain valuable information on the bulk gauge bosons and fermions.

This paper is organized as follows. In Sec II, we briefly  review the set up of the RS model. 
In Sec III, we explain that typical corrections to the Higgs couplings are consistent with the current  constraints. 
In Sec IV, we study the KK mass dependence  of the Higgs coupling corrections.
In Sec V, we clarify the source of the coupling corrections for each channel.
In Sec VI, we select the model points and show how Higgs coupling measurements
separately determines the KK contributions and Higgs-radion mixing. 
Sec VII is devoted to the conclusions.

\section{Randall-Sundrum model}
The RS model can resolve the gauge hierarchy problem by one $AdS$ extra dimension
since this model replaces the gauge hierarchy by an exponential function of  the fifth dimensional curvature $k$ times the size of the fifth dimension $r_c\pi$ which is called the warp factor $e^{-kr_c\phi}$~\cite{Randall:1999ee}.
The background metric of the original RS model is given by
\begin{equation}
ds^2=e^{-2kr_c\phi}\eta_{\mu \nu}dx^\mu dx^\nu+r_c^2 d\phi^2.
\end{equation}	
The fifth dimension has the topology $S_1/Z_2$ and is parametrized by $\phi$ $(0<\phi<\pi)$.
The boundaries $\phi=0,\pi$ are locations of the four dimensional hidden ($\phi=0$) and visible ($\phi =\pi$) branes.
The four dimensional effective action  is obtained by integrating out the extra dimension. 
The fundamental mass parameter in the visible brane is reduced from the Planck scale  $m_{\rm{pl}}$ by the warp factor after canonical normalization of the kinetic term.
The electroweak scale is obtained for $kr_c\pi \sim 30$, when $m_{\rm{pl}}e^{-kr_c\phi} \sim $ TeV.

The value of  $r_c$, which is the size of the extra dimension is not determined 
if the distance between the visible and hidden branes is not stabilized . 
A mechanism for stabilizing the size of the extra dimension called Goldberger-Wise mechanism is proposed in~\cite{Goldberger:1999uk,Goldberger:1999un}.
In this mechanism,  a massive five-dimensional scalar field with self-interaction terms at the two branes
generates a potential for the radion in the effective theory. 
The radion corresponds to a fluctuation of the fifth dimension size.
An appropriate value $r_c$ to realize the TeV scale at the visible brane may be achieved without large fine tuning among the parameters of the scalar potential when the radion mass can also be below 1 TeV~\cite{Goldberger:1999un,Csaki:2000zn}.

The five dimensional metric may be expanded 
by a radion field $r(x)$ as follows
\begin{equation}
ds^2=e^{-2(kr_c\phi+F(x,\phi))}\eta_{\mu \nu}dx^\mu dx^\nu-(1+2F(x,\phi))^2r_c^2 d\phi^2,
\end{equation}
where $F(x,\phi)=r(x)R(\phi)$, and  $R(\phi)$ is determined by the Einstein equation.
In the limit that the back-reaction can be ignored, $F(x,\phi)$ is given by
\begin{equation}
F(x,\phi)=\frac{r(x)}{\Lambda_\phi}e^{2kr_c(\phi-\pi)},
\end{equation}	 
where $\Lambda_\phi$ is the vacuum expectation value of the radion $\Lambda_\phi=\sqrt{6}M_{pl}e^{-kr_c\pi}$. 
$\Lambda_\phi$ is almost the same as the lightest KK particle mass. 
The lower bound of $\Lambda_\phi$ is around 10 TeV to satisfy  the kaon mixing constraint.

The radion couples to the trace of the SM energy-momentum tensor.
The interactions of the radion with the SM fermions and the massive gauge bosons are similar to that of the Higgs boson, but these are suppressed by $\Lambda_\phi \geq 10$ TeV instead of  the Higgs boson vacuum expectation value $v=246$ GeV.  
However, quantum corrections generate the effective radion coupling to gluons and photons, which are known as the trace anomaly terms $\mathcal{L}_{\rm{anom}}$~\cite{Csaki:2000zn}.

Next we consider the bulk matter fermions in the RS model. 
In this model, we can also relax hierarchies among the SM Yukawa couplings by profiles of the fifth dimensional fermion wave functions~\cite{Grossman:1999ra,Gherghetta:2000qt}.
 The profile is controlled by 
the five dimensional masses of the bulk fermions, $m^{5D}_{L,R}$. 
The four dimensional Yukawa couplings are determined by overlapping between the fifth dimensional wavefunction and  the Higgs boson at the visible brane.
The large hierarchy among the SM Yukawa couplings can be derived 
without fine tuning among $\vert c_{L,R}\vert $, where  $c_{L,R}=m^{5D}_{L,R}/k<1$. 
Namely, we can take model parameters where the fifth dimensional wavefunctions of the third generation fermions 
 are localized near the visible brane
and those of the first and second generations are localized near the hidden brane. 
Since the overlap between the fifth dimensional wavefunction of  the third generation fermions and the Higgs boson
is large, the effective Yukawa coupling of the third generation fermions are also large.
Conversely, in the case of the first and second generations, the four dimensional Yukawa couplings are small due to the suppressed wavefunction at the visible brane, even though the five dimensional Yukawa couplings can be large. 

When the SM fermions and gauge bosons propagate in the fifth dimension,
KK modes of each field contribute to the Higgs(radion)-$\gamma\gamma$ and -$gg$ couplings through triangle loop diagrams. 
We call this contribution $\mathcal{L}_{\rm{triangle}}$.
The contribution of the KK $W$ boson is determined by the profiles in the fifth dimension which is a function of  $\Lambda_\phi$.
The KK fermion contributions to the same vertex are determined  by $\Lambda_\phi$ and  $c_{L,R}$.
Due to the $c_{L,R}$ dependence, the KK fermion contributions can vary. 
The couplings of the KK-top to the Higgs boson and radion are the largest among the KK fermions, 
however, the KK modes of light fermions may also contribute significantly by choosing profiles of the left-handed and right-handed KK modes separately. 
We call this contribution $\mathcal{L}_{\rm{tree}}$, which is inversely proportional to the fifth dimensional volume and important for massless gauge bosons.
The interaction terms of the radion with the SM massless gauge bosons
are given by
\begin{align}
\mathcal{L}_{\rm{tree}}&=-\frac{r(x)}{4\Lambda_\phi kr_c\pi}(F^a_{\mu \nu}F^{a\mu \nu}+F_{\mu \nu}F^{\mu \nu}), \\[6pt]
\mathcal{L}_{\rm{anom}}+\mathcal{L}_{\rm{triangle}}&=-\frac{r(x)}{4\Lambda_\phi}\Big(\frac{b_{QCD}^{r}\alpha_s}{2\pi}F^a_{\mu \nu}F^{a\mu \nu}+\frac{b_{EM}^{r}\alpha}{2\pi}F_{\mu \nu}F^{\mu \nu}\Big), \\[6pt]
\mathcal{L}_{r}^{\gamma \gamma,gg}&=\mathcal{L}_{\rm{triangle}}+\mathcal{L}_{\rm{tree}}+\mathcal{L}_{\rm{anom}} \\[6pt]
&=-\frac{r(x)}{4\Lambda_\phi}\Big[\Big(\frac{1}{kr_c\pi}+\frac{\alpha_s}{2\pi}b_{QCD}^{r}\Big)F^a_{\mu \nu}F^{a\mu \nu}+\Big(\frac{1}{kr_c\pi}+\frac{\alpha}{2\pi}b_{EM}^{r}\Big)F_{\mu \nu}F^{\mu \nu}\Big],
\end{align}\label{Eq:radiongp }
where
\begin{align}
b_{QCD}^{r}&=7+(F_f+F_f^{KK}) \label{QCDanomaly},\\[6pt]
b_{EM}^{r}&=-11/3+8/3(F_f+F_f^{KK})-(F_W+F_W^{KK}), \\[6pt]
F_f&=\tau_f(1+(1-\tau_f)f(\tau_f)) \\[6pt]
F_W&=2+3\tau_W+3\tau_W(2-\tau_W)f(\tau_W), \\[6pt]
f(\tau)&=(Arc\sin\frac{1}{\sqrt{\tau}})^2  \ \rm{for} \ \tau > 1, \\[6pt]
f(\tau)&=-\frac{1}{4}(\log\frac{\eta_+}{\eta_-}-i\pi)^2 \ \rm{for} \ \tau < 1,  
\end{align}
and $\tau_i=\Big(\frac{2m_i}{m_r} \Big)^2$, $\eta_{\pm}=1\pm \sqrt{1-\tau} $ and 
$m_i$ is the mass of the particle involved in the triangle loop diagram.
$F_f^{KK}$ ($F_W^{KK}$) are the sum of the contributions of the KK fermions (KK $W$ bosons) up to higher modes.

The interaction terms between the Higgs boson and massless gauge bosons are given by
\begin{equation}
\mathcal{L}_{h}^{\gamma \gamma,gg}=-\frac{h(x)}{4v}\Big[\frac{\alpha_s}{2\pi}b_{QCD}^{h}F^a_{\mu \nu}F^{a\mu \nu}+\frac{\alpha}{2\pi}b_{EM}^{h}F_{\mu \nu}F^{\mu \nu}\Big],
\end{equation}
where $b_{QCD}^{h}=F_f+F_f^{KK}$, and $b_{EM}^{h}=8/3(F_f+F_f^{KK})-(F_W+F_W^{KK})$.
There is no term corresponding to $\mathcal{L}_{\rm{tree}}$ and $\mathcal{L}_{\rm{anom}}$
because the Higgs boson lives on  the visible brane in our set up.

In addition  we need to  consider the Higgs-radion mixing in the effective four dimensional theory~\cite{Dominici:2002jv}.
The mixing is induced by a curvature-higgs mixing term in the four dimensional effective action.
\begin{equation}
\mathcal{L}_\xi=\sqrt{g_{\rm{ind}}}\xi R(g_{\rm{ind}})H^\dag H,
\end{equation}
where $\xi$ is a mixing parameter, $g_{\rm{ind}}$ is the four dimensional induced metric $e^{-2(kr_c\pi+\frac{r(x)}{\Lambda_\phi})}\eta_{\mu \nu}$, and $R(g_{\rm{ind}})$ is the four dimensional Ricci scalar of the induced metric.

The relation between the $r$, $h$ and the mass eigenstates $h_m$, $r_m$ is given by
\begin{align}
h&=(\cos \theta-\frac{6\xi \gamma}{Z}\sin \theta)h_m+(\sin \theta+\frac{6\xi \gamma}{Z}\cos\theta) r_m
\equiv d h_m+cr_m, \label{mixing} \\
r&=\cos\theta\frac{r_m}{Z}-\sin\theta \frac{h_m}{Z}
\equiv ar_m+bh_m, \\
\tan2 \theta&=12\xi \gamma Z \frac{m_h^2}{m_r^2-m_h^2(Z^2-36\xi^2 \gamma^2)},
\end{align}
where $\gamma=\frac{v}{\Lambda_\phi}$, and $Z^2=1+6\xi \gamma^2(1-6\xi)$.
The mass eigenvalues of the mixed scalars are given by
\begin{equation}
m_{\pm}^2=\frac{1}{Z^2}\Big(m_r^2 +\beta m_h^2\pm \sqrt{(m_r^2 +\beta m_h^2)^2-4Z^2 m_r^2 m_h^2}\Big),
\end{equation}
where $\beta=1+6\xi \gamma^2$. 
The factor $Z$ arises from canonical normalization of the radion kinetic term, 
leading to anomalously large interactions when $Z\rightarrow 0$. 
In this paper, we consider the parameters where 
 $h_m$ is the SM-like Higgs boson and $m_h=125$ GeV. 
On the other hand, $r_m$ interactions with the SM particles are small enough so as to be consistent with the recent Higgs searches at the ATLAS and CMS.  

The couplings of $h_m$ and $r_m$ to the SM particles are affected by the Higgs-radion mixing.
The interactions with the SM massive gauge bosons are given by
\begin{align}
 \mathcal{L}_{h_m(\xi \neq 0)}^{WW,ZZ}&=-(d+\gamma b)\frac{h_m(x)}{v}\Big[2M_W^2W_\mu^{(0)+}W^{(0)\mu -}+M_Z^2Z_\mu^{(0)}Z^{(0)}\Big], \\
\mathcal{L}_{r_m(\xi \neq 0)}^{WW,ZZ}&=-(c+\gamma a)\frac{r_m(x)}{v}\Big[2M_W^2W_\mu^{(0)+}W^{(0)\mu -}+M_Z^2Z_\mu^{(0)}Z^{(0)}\Big].   
\end{align} 
Similarly, the interactions with  the SM fermions are given by
\begin{align}
\mathcal{L}_{h_m(\xi \neq 0)}^{ff}&=(d+\gamma b)\frac{h_m(x)}{v}m_f\bar{\psi}^{(0)}\psi^{(0)}, \\ 
\mathcal{L}_{r_m(\xi \neq 0)}^{ff}&=(c+\gamma a)\frac{r_m(x)}{v}m_f\bar{\psi}^{(0)}\psi^{(0)}, 
\end{align}
and the interaction with the SM massless gauge bosons are as follows
\begin{align}
\mathcal{L}_{h_m(\xi \neq 0)}^{\gamma \gamma,gg}&=-\frac{h_m(x)}{4v}\Big[\Big \{\gamma b(\frac{1}{kr_c\pi}+\frac{\alpha_s}{2\pi}b_{QCD}^{r}\Big)+d \frac{\alpha_s}{2\pi}b_{QCD}^{h} \Big \}F^a_{\mu \nu}F^{a\mu \nu} \\ \nonumber
&+ \{\gamma b(\frac{1}{kr_c\pi}+\frac{\alpha_s}{2\pi}b_{EM}^{r}\Big)+d \frac{\alpha_s}{2\pi}b_{EM}^{h} \Big \}F_{\mu \nu}F^{\mu \nu}\Big],  \\ 
\mathcal{L}_{r_m(\xi \neq 0)}^{\gamma \gamma,gg}&=-\frac{r_m(x)}{4v}\Big[\Big \{ \gamma a(\frac{1}{kr_c\pi}+\frac{\alpha_s}{2\pi}b_{QCD}^{r}\Big)+c \frac{\alpha_s}{2\pi}b_{QCD}^{h} \Big \}F^a_{\mu \nu}F^{a\mu \nu} \\ 
&+ \{\gamma a(\frac{1}{kr_c\pi}+\frac{\alpha_s}{2\pi}b_{EM}^{r}\Big)+ c \frac{\alpha_s}{2\pi}b_{EM}^{h} \Big \}F_{\mu \nu}F^{\mu \nu}\Big]. \nonumber
\end{align}

\section{Higgs coupling measurements at the LHC and the ILC}
Following the discovery of the 125 GeV Higgs boson at the LHC, 
future collider projects now focus on the measurements of the 
Higgs couplings.  
BSM scenarios often predict a non-minimal Higgs sector. 
The Higgs boson production cross section and branching ratios may deviate from the 
SM predictions. 
The sign of new physics may be obtained from the coupling measurements at the future colliders.  
Especially, we can not discover the KK particles of the RS model even at the 14 TeV LHC if we should take into account the current  FCNC constraint to the KK particle masses seriously. 
Measurements of the Higgs effective couplings are very important to prove this model.

In this paper, the deviations of couplings $g(HAA)$ from the SM predictions $g(HAA)_{SM}$ are expressed by $d(A)$ which is defined as
\begin{equation}
\frac{g(HAA)}{g(HAA)_{\rm SM}}=1+d(A),
\end{equation}
where $A=f,W,Z,\gamma,g$. 
In~\cite{Peskin:2012we,Peskin:2013xra,CMSpre}, the precisions of coupling measurements at the 14 TeV LHC and the ILC are estimated.

LHC is a $pp$ collider. Although the Higgs boson production cross section from the 
gluon fusion is large, the uncertainty in the higher order QCD correction
is not small. In addition, the accessible channels suffer from large QCD backgrounds. 

The detection ratios of the Higgs boson is an important quantity to express the Higgs signal strength at the LHC. 
The 125 GeV Higgs bosons are mainly generated by a gluon fusion process at the LHC, and  the observed number of events is proportional to the cross section times the branching ratio in each channel.
To compare the data of the Higgs boson searches at the LHC with the theoretical predictions, 
detection ratio of the Higgs-like mass eigenstate $H$ normalized by the SM predictions is defined as 
  \begin{equation}
DR_{H}(X) \equiv \frac{\Gamma_{H} (gg)BR(H \rightarrow X)}
{\Gamma_{H_{\rm SM}}(gg)BR(H_{\rm SM} \rightarrow X)}
\end{equation}
where $X$ is $ff$, $WW$, $ZZ$, and $\gamma \gamma$. 
Some of the relevant signal strengths are listed in Table \ref{ATLASCMSHiggs}.
The error is currently around $30\%$, which would be reduced down to $7\sim8\%$ at the HL-LHC  ($\sqrt{s}=14$~TeV and $\int dt L = 3000$ fb$^{-1}$), and may be reduced to $3\sim4\%$ if the theoretical errors can be reduced by half~\cite{Peskin:2013xra}.

\begin{table}
\begin{center}
\begin{tabular}{|l||c|c|c| }
\hline 
mode & ATLAS\cite{ATLAStalk} & CMS\cite{CMStalk}& HL-LHC\cr 
\hline
$H\rightarrow W^+W^-$ & $1.0 \pm0.3$ &  $0.68\pm 0.20$ & $\pm0.07(0.04)$ \cr 
$H\rightarrow Z^0Z^0$  & $1.5\pm 0.4$  &      $0.92\pm 0.20$& $\pm0.07(0.03)$\cr 
$H\rightarrow \gamma \gamma$ & $1.6 \pm 0.3$ &  $0.77\pm 0.27$& $\pm0.08(0.04)$ \cr
\hline
\end{tabular}
\end{center}
\caption{The latest result of the SM Higgs boson signal strengths measured by CMS and ATLAS.
For the HL-LHC(CMS 3000 fb$^{-1}$) the numbers in parentheses assume the reduction of theoretical errors by half.}
\label{ATLASCMSHiggs}
\end{table}

On the other hand, the proposed International Linear Collider(ILC) at $\sqrt{s}=250$~GeV 
to 1000~GeV and with an integrated luminosity of up to 2500 fb$^{-1}$ can measure the Higgs couplings even more precisely (Table \ref{Higgscouplingprecisions}). The merit of the ILC over the LHC is its clean environment. The production cross sections are calculated with good precision, because these are EW processes.  Because QCD background at the ILC is smaller, the branching ratios into $H\rightarrow b\bar{b}, c\bar{c}$ are accessible with a high precision.

\begin{table}[ht]\label{tb:peskin}
\begin{tabular}{c|ccccccccc} 
 & LHC &  HL-LHC & ILC250 & 500 & 500up & 1000 & 1000up  \\ \hline
channel& $300 fb^{-1}$ & $3000 fb^{-1}$ &250 fb$^{-1}$ &500 fb$^{-1}$ &1.6 ab$^{-1}$ & 1 ab$^{-1}$ & 2.5 ab$^{-1}$  \\ \hline
$W$ & 4.6(3.5)&2.3(1.2) &4.6(1.4) &0.46(0.43) &0.22(0.21) &0.19(0.19) &0.15(0.15)  \\ 
$g$ & 6.3(4.1)& 4.4(2.0)&6.1(2.0) &2.0(1.4) &0.96(0.85) &0.79(0.72) &0.60(0.56)  \\ 
$b$ & 10.2(7.6)& 6.0(2.9)&4.7(1.8) &0.97(0.80) &0.46(0.43) &0.39(0.37) &0.32(0.29)  \\ 
$Z$ & 6.6(5.0)&4.3(2.2) &0.78(0.57) &0.50(0.47) &0.23(0.23) &0.22(0.22) &0.22(0.21)  \\ 
$\gamma$ &7.3(4.3) &5.3(2.4) &18.8(2.0) &8.6(1.8) &4.0(1.7) &2.9(1.6) &1.9(0.83)  \\ \hline
\end{tabular}
\caption{Higgs coupling precisions $\delta d(A)$ in [\%] as estimated by~\cite{Peskin:2013xra} at each stage of the ILC and LHC. Numbers in parentheses of $W,g,b,Z$ channels are the combinations of the ILC data set and the optimistic CMS estimates(Scenario 2~\cite{Peskin:2013xra}).  Numbers in parentheses of $\gamma$ channel are the combinations of the ILC data and CMS Scenario 2 set with $\delta(BR(\gamma\gamma)/BR(ZZ))=3.6\%$.}
\label{Higgscouplingprecisions}
\end{table}

Table~\ref{Higgscouplingprecisions} shows the errors in coupling measurements $\delta d(A)$ at the 14 TeV LHC and the ILC250, 500, 1000 as estimated in~\cite{Peskin:2012we}. 
For the LHC at 300 fb$^{-1}$, the precisions of $g(HWW),g(HZZ),g(Hgg)$ and $g(H\gamma\gamma)$ are at the $3 \sim 5 \%$ level in Scenario 2. The precisions of $g(Hb\bar{b})$ is much worse about $8 \%$.
At the ILC250, the precision of the each channel is improved except for the $t\bar{t}$ channel. 
The precision of the $g(HZZ)$ is below $1\%$ level.
Moreover, at the ILC500 and ILC1000, the precisions of the $g(Ht\bar{t}), g(HWW)$ and $g(Hb\bar{b})$ are below $1\%$ level.
Although it is not listed in the table, for $g(Ht\bar{t})$ it is about $4\%$.

While  most of the couplings would be measured with impressive accuracy at the ILC, the precision of  $Br(H\rightarrow \gamma \gamma)$ is somewhat disappointing. 
It is around 4\%  at the ILC 500up, which is only a factor 2 improvement 
from the LHC $300 fb^{-1}$. 
This is due to the rather small cross section of the Higgs boson production at the ILC, $\sigma(e^+e^-\rightarrow HZ) \sim 200$ fb, compared with $\sigma( gg\rightarrow H) \sim 50$~pb at the LHC.  

However, the recent studies shows that the ratio of the branching fractions 
$Br(H\rightarrow \gamma \gamma) /Br( H\rightarrow Z Z)$ can be measured rather 
precisely at the HL-LHC. 
The merit of taking the ratio is that the uncertainty of  the production 
cross section in $DR_{H}(\gamma\gamma)$ and $DR_{H}(ZZ)$ cancels out in the ratios. 
Moreover, both ATLAS and CMS plan to keep the  trigger threshold of $\gamma, e, \mu$  close to the current level by the detector upgrades so that signal efficiency would not reduce significantly. 
In this paper, we take the error $\delta(Br(\gamma\gamma)/Br(ZZ))=3.6\%$~\cite{Peskin:2013xra}.
The numbers after including this constraint are listed in Table~\ref{Higgscouplingprecisions} in parentheses.

\section{Coupling deviations in the RS model at $\Lambda_{\phi}=10$~TeV}
In the RS model, the deviations of the fermions, $W$ boson and $Z$ boson couplings to the 125 GeV scalar boson from the SM are common.
It occurs due to the Higgs-radion mixing. 
The ratio of the couplings are expected as $1+d(f,W,Z)=d+\gamma b$, where $d$, $\gamma b$ are defined in Eq~\ref{mixing}.
Therefore, we use $d(f,W,Z)$, $d(\gamma)$ and $d(g)$ to express the deviations of the couplings from the SM.

On the other hand, 
$d(g)$ and $d(\gamma)$ are loop level quantities. 
There are contributions of the KK particles in addition to the Higgs-radion mixing.
The contributions of the KK quarks decrease $d(g)$ and increase $d(\gamma)$.
The KK leptons and KK $W$ bosons also contribute to the $d(\gamma)$.
The effects of the Higgs-radion mixing and the KK loops are  suppressed by  $\Lambda_\phi$ and the KK particle masses.
However, the ILC can measure the couplings of the 125 GeV scalar boson to the SM particles very precisely as  discussed in Section III. 
Therefore, we expect to measure the new physics effects at the LHC and the ILC.

The quantity we measure directly at the LHC is the detection ratio.
The deviations of the detection ratios of the fermions, $W$ boson and $Z$ boson from the SM predictions are common as the  $d(A)$ case and depend on the Higgs-radion mixing and $F_q^{KK}$.
We call it as  $DR_{h_m}(ff,WW,ZZ)$.
The measurement of $DR_{h_m}(\gamma\gamma)$ gives independent information because it also depends on loop effects of the KK modes of fermions and gauge bosons. 
Currently, there is relatively large positive deviation from the SM prediction in the $\gamma\gamma$ detection ratio in the ATLAS, while that in the CMS is smaller than the SM prediction.

\begin{figure}[t]
 \begin{minipage}{0.45\hsize}
  \begin{center}
   \includegraphics[bb=7 0 551 409, width=70mm]{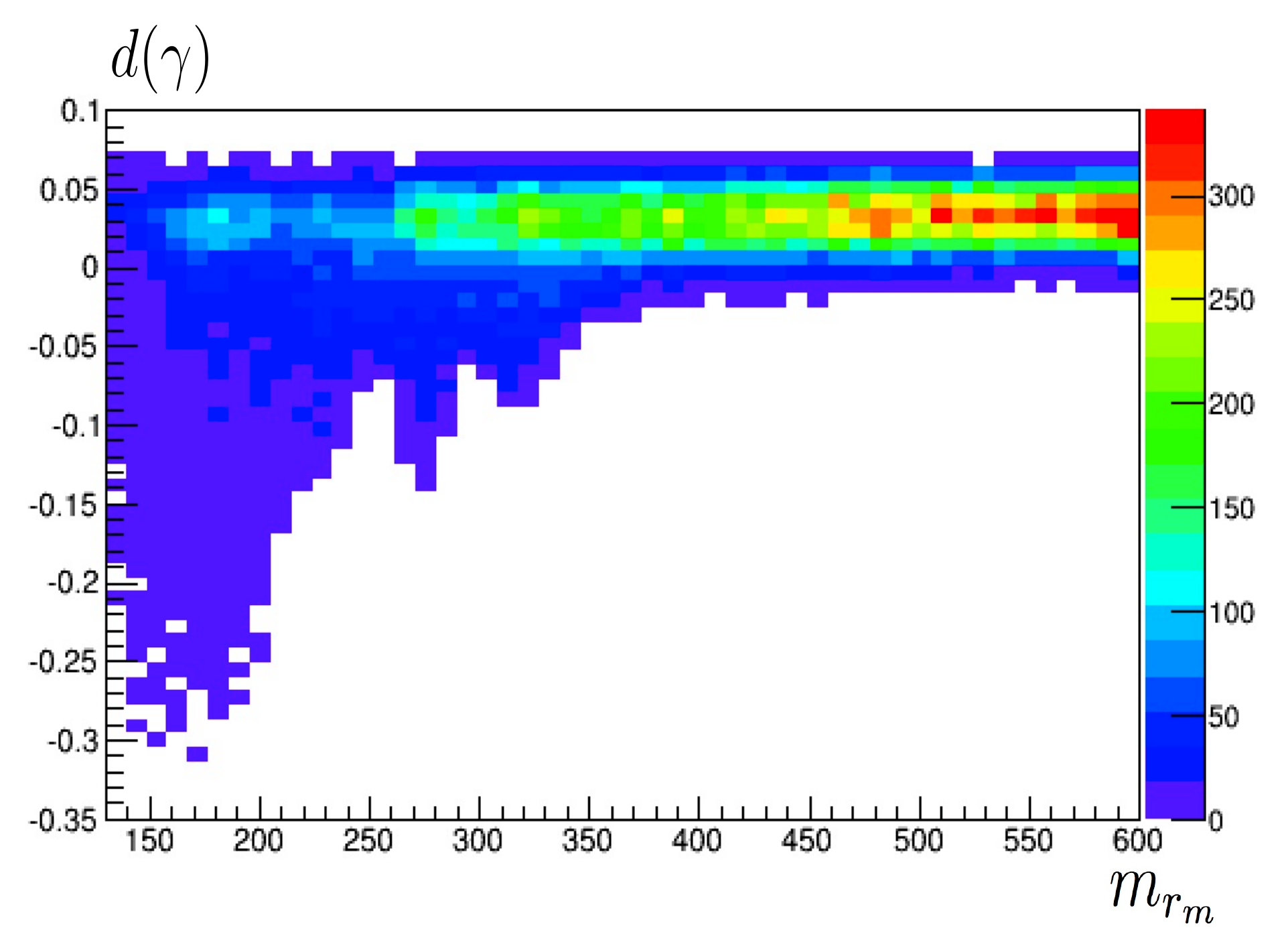}
   (a)
  \end{center}
 \end{minipage}
 \begin{minipage}{0.45\hsize}
  \begin{center}
   \includegraphics[bb=7 0 579 395, width=70mm]{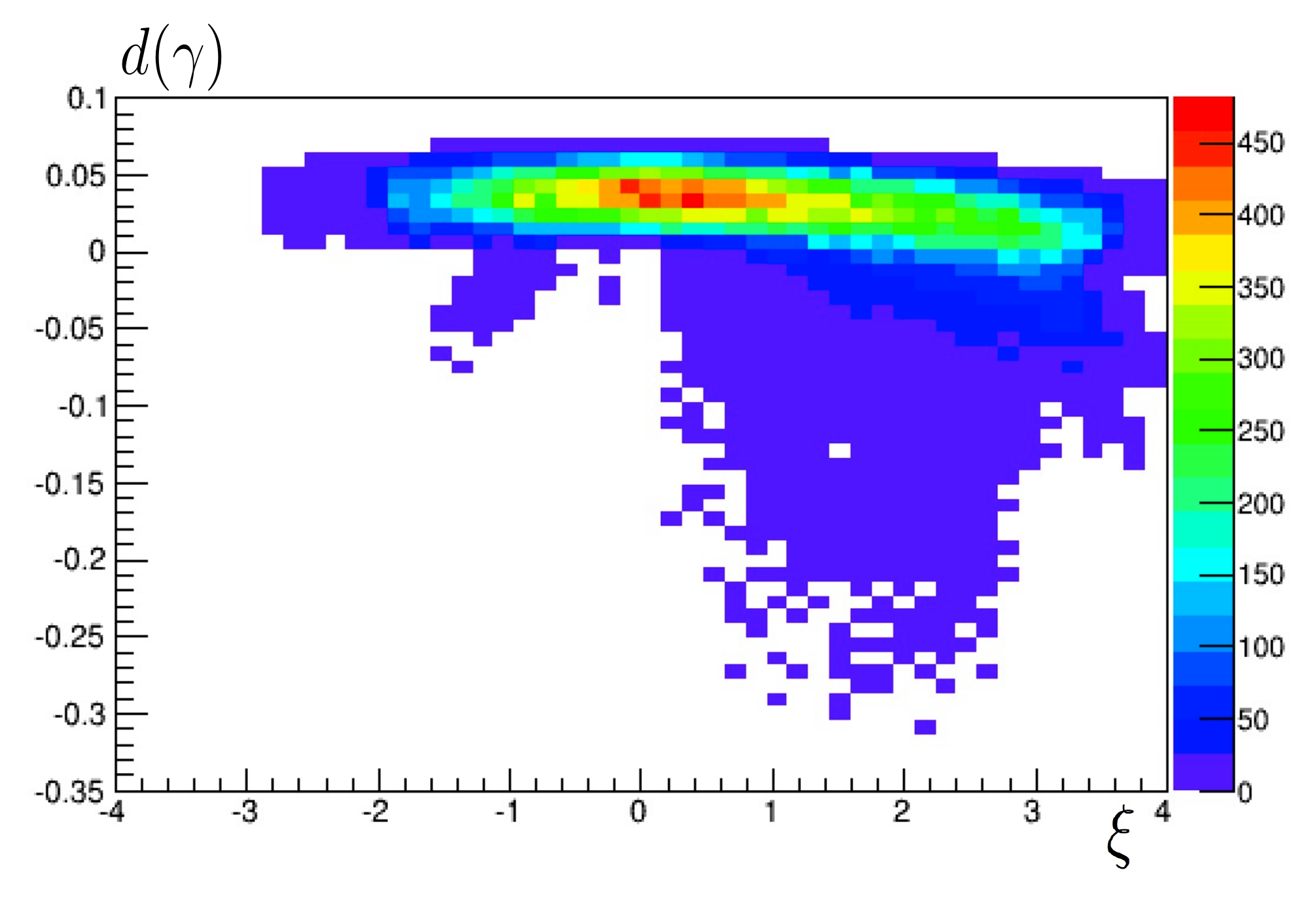}
   (b)
  \end{center}
 \end{minipage}
 \caption{(a) The distribution of the model points in $d(\gamma)$ vs $m_r$ plane and (b) $d(\gamma)$ vs $\xi$ plane at $\Lambda_\phi$=10 TeV.}
\label{fig:one}
\end{figure}

\begin{figure}[tbph]
 \begin{minipage}{0.45\hsize}
  \begin{center}
   \includegraphics[bb=7 0 583 403, width=75mm]{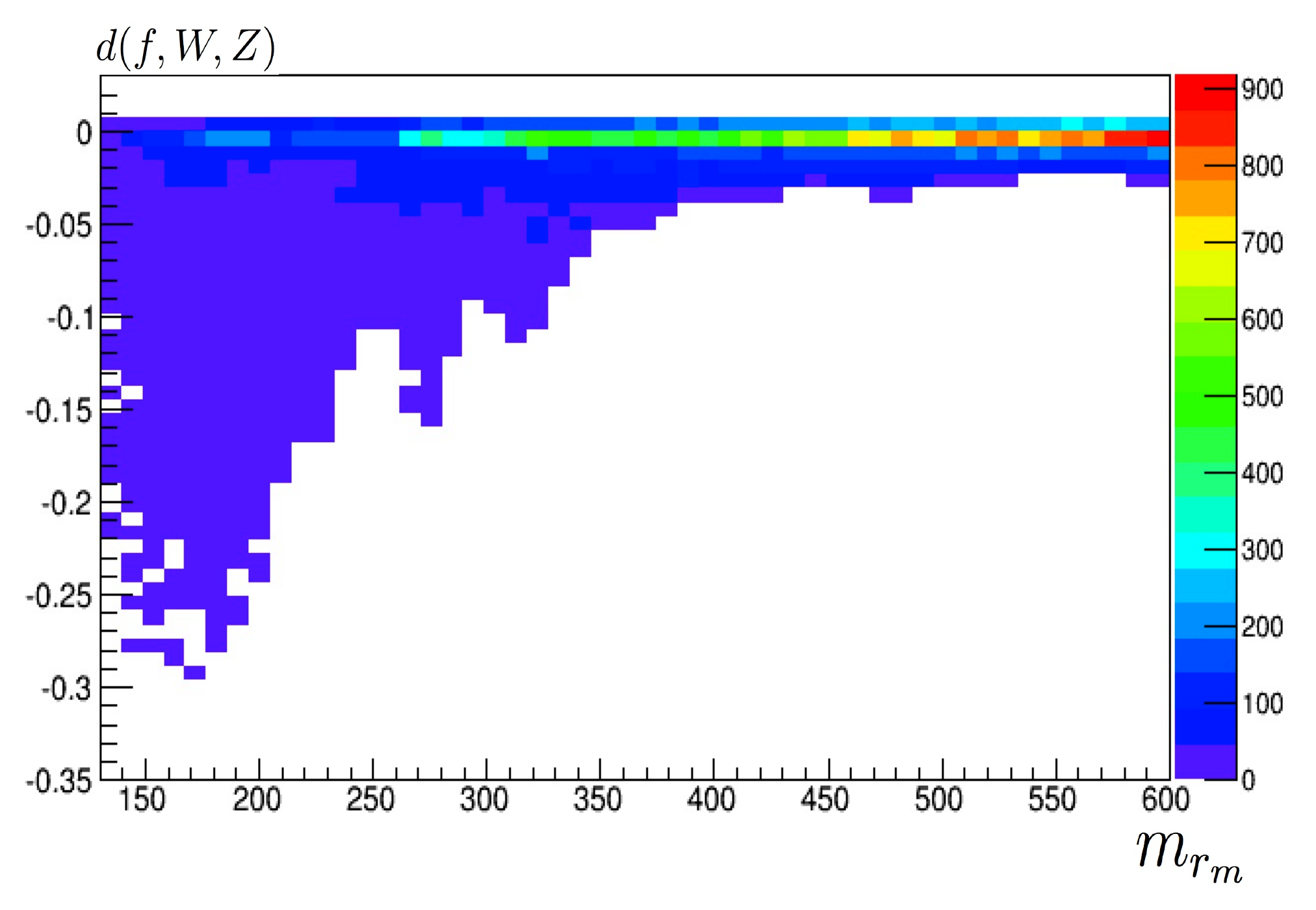}
   (a)
  \end{center}
 \end{minipage}
 \begin{minipage}{0.45\hsize}
  \begin{center}
   \includegraphics[bb=0 0 597 399, width=75mm]{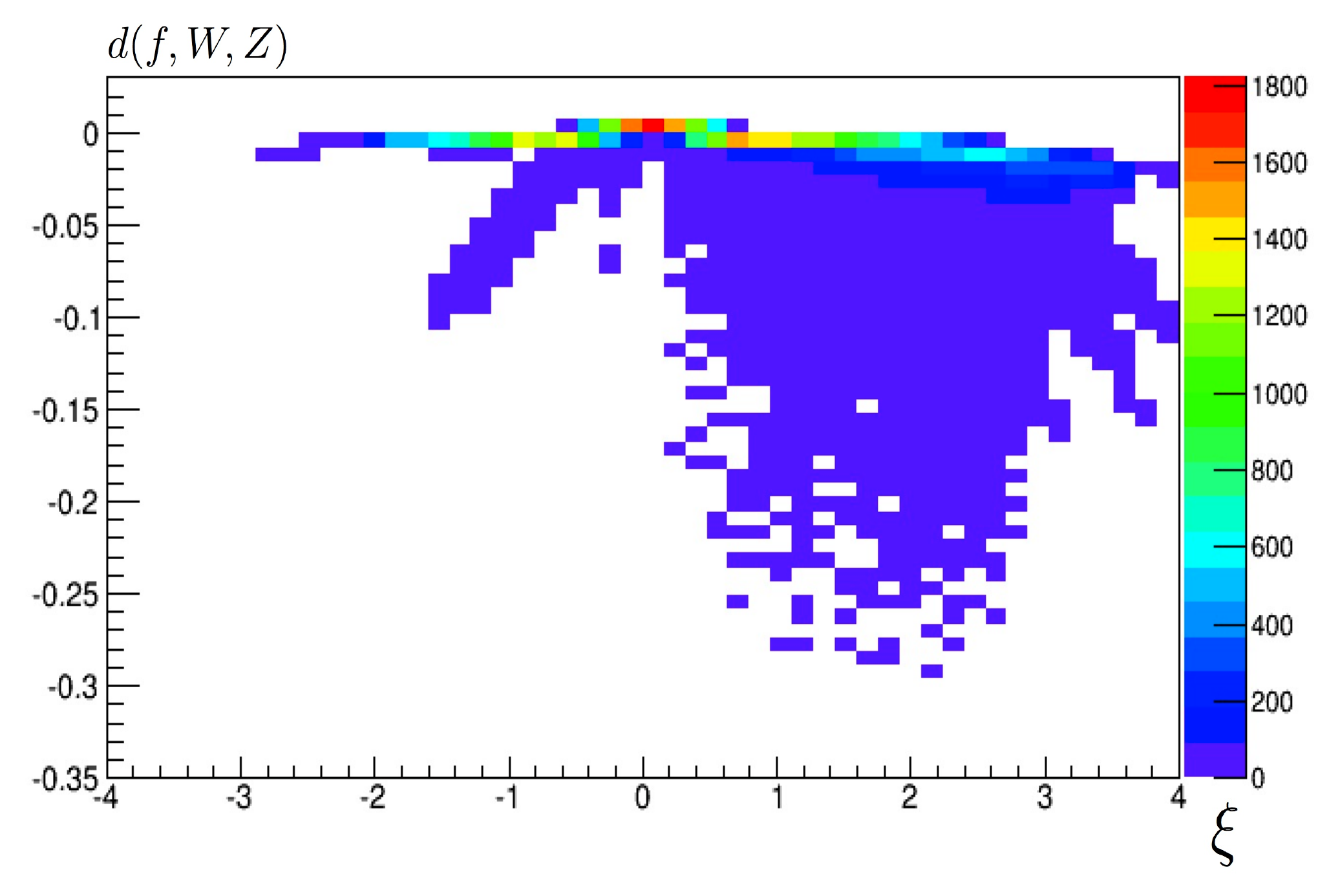}
   (b)
  \end{center}
 \end{minipage}
  \caption{(a) The distribution of the model points in $d(f,W,Z)$ vs $m_r$ plane and (b) $d(f,W,Z)$ vs $\xi$ plane at $\Lambda_\phi$=10 TeV. }
 \label{fig:two}
\end{figure}

\begin{figure}[tbph]
 \begin{minipage}{0.45\hsize}
  \begin{center}
   \includegraphics[bb= 8 0 678 406, width=70mm ]{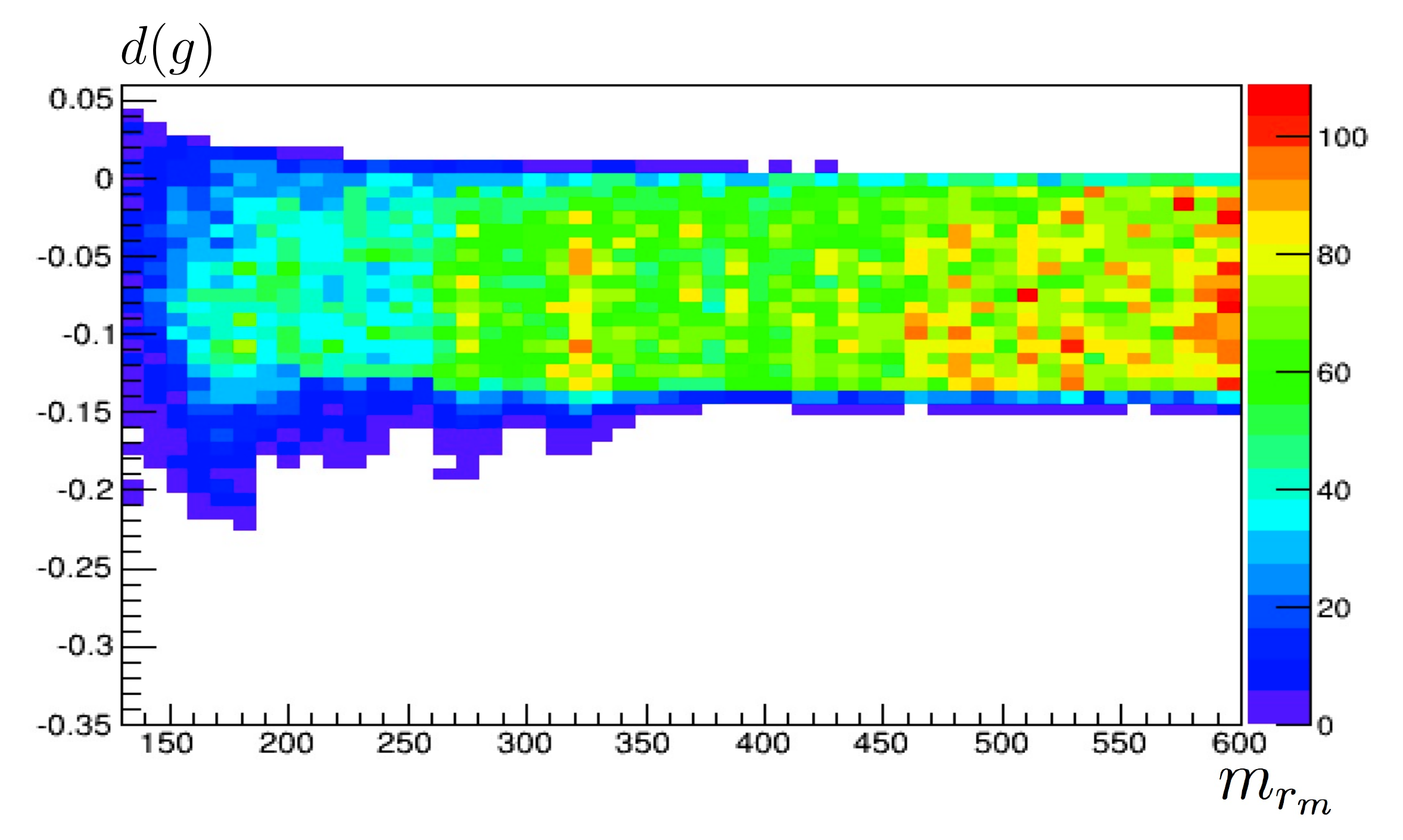}
   (a)
  \end{center}
 \end{minipage}
 \begin{minipage}{0.45\hsize}
  \begin{center}
   \includegraphics[bb= 8 0 667 399, width=70mm]{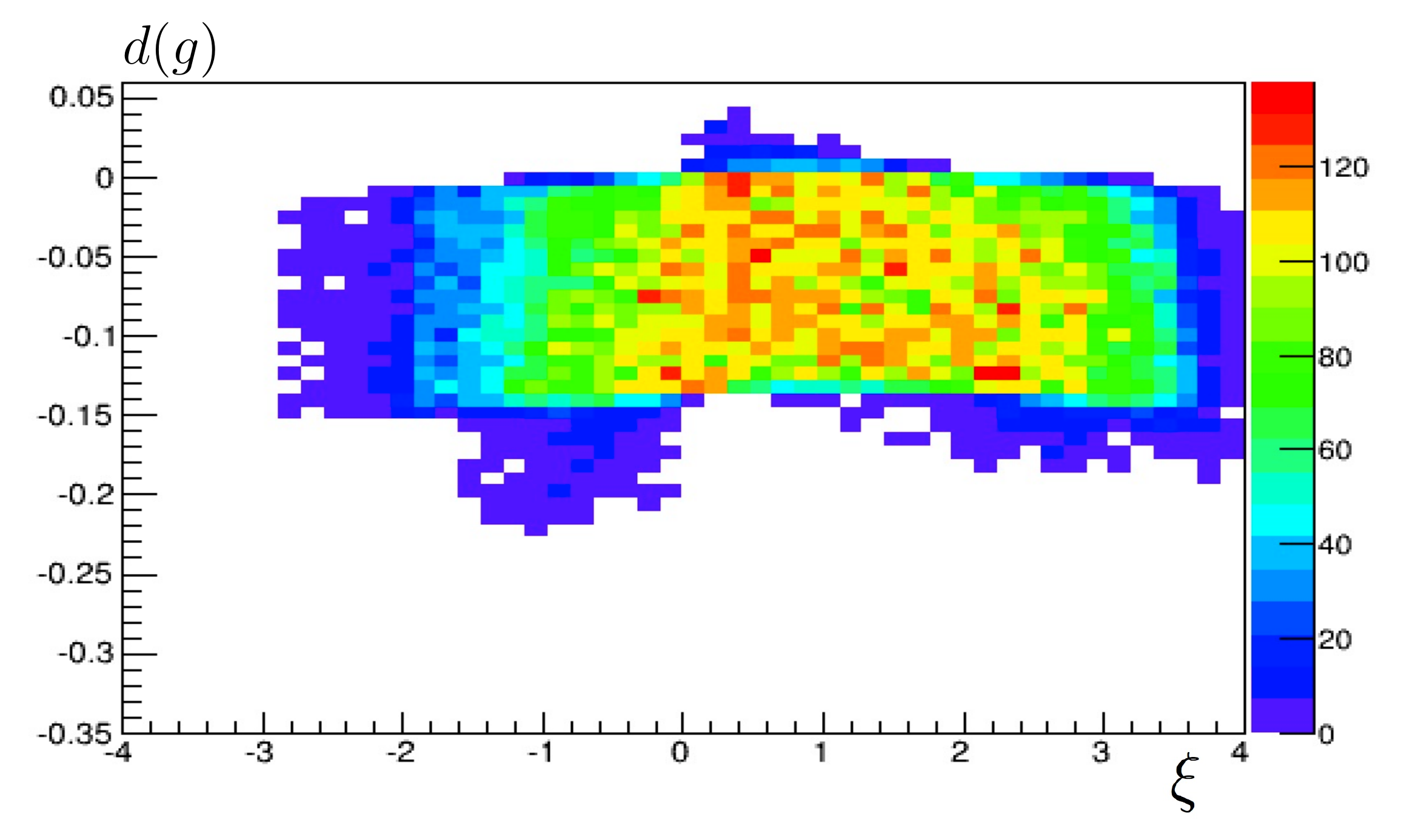}
   (b)
  \end{center}
 \end{minipage}
 \caption{(a) The distribution of model points in $d(g)$ vs $m_r$ plane, and (b) $d(g)$ vs $\xi$ plane at $\Lambda_\phi$=10 TeV.}
\label{fig:three}
\end{figure}

From now on, we show results of a parameter scan in the RS model.
We generate model points which have uniform distribution between 130 GeV$<m_{r_m}<600$ GeV, 
$\xi_{\rm{min}}<\xi<\xi_{\rm{max}}$, $0<F^{KK}_t<0.09$, and $0<F^{KK}_l<0.083$.
Here, $\xi_{\rm{max}}$ and $\xi_{\rm{min}}$ are theoretical lower and upper bounds.
We also fix $\Lambda_\phi$=10 TeV and $DR_{h_m}(\gamma\gamma)>0.7$. 
Fig.~\ref{fig:one}, \ref{fig:two} and \ref{fig:three} show distribution of model points $d(\gamma)$, $d(f,W,Z)$ and $d(g)$ vs (a) mass of radion-like state $m_{r_m}$ and (b) the mixing parameter $\xi$ respectively.
In each figure, the color code  represents  the  number of model points in the same bin.  
At the LHC, no other scalar boson for the 125~GeV Higgs boson has been discovered. Therefore, if the detection ratio of the mixed radion $r_m$ exceeds the 95$\%$ CL upper limit of the ATLAS and CMS Higgs searches~\cite{atlasconstraint,Chatrchyan:2013mxa}, we 
exclude those points from the figures. 
The discovery potential of $r_m$ at ILC will be discussed in section VIII.

The maximal value of $d(\gamma)$ is about 0.06 in Fig.~\ref{fig:one} and it is achieved for the small Higgs-radion mixing region ($\xi\sim 0$).
Most of the model points have $\vert d(\gamma)\vert > 0.01$, and the deviation may be detected if the ultimate sensitivity of $\delta d(\gamma)=0.83$ \% in Table~\ref{Higgscouplingprecisions} is achieved. 
The Higgs-radion mixing suppresses the coupling of $h_{m}$ to the SM particles. This can be seen in the $d(f,W,Z)$ distribution in Fig. \ref{fig:two} where model points with large negative deviation are found for $\vert \xi\vert \gg0$  and $m_{r_m}<250$ GeV. 
In Fig.~\ref{fig:one}, \ref{fig:two}, and \ref{fig:three} the lower limits of $d(\gamma)$, $d(f,W,Z)$, and $d(g)$ are constrained by the condition that $r_m$ is not discovered by the current Higgs boson searches.
In Fig.~\ref{fig:three}, we see that the 
KK quark contribution to $d(g)$ is negative except at some exceptional points where $m_{r_m}$ is close to $m_{h_m}$.  Enhancement of $d(g)$ for small $m_{r_m}$ is due to the trace anomaly contribution of $g(rgg)$ through the Higgs-radion mixing. 
Indeed, the enhancement does not occur at $\xi=0$, where the mixing is zero, and for very large $|\xi|$  where the mixing reduces the Yukawa couplings significantly.
Again, the deviation of the couplings is larger than the ultimate  coupling sensitivity listed in Table.~\ref{Higgscouplingprecisions}.

\section{ Maximal deviation and the new physics scale}

\begin{figure}[htbp]
 \begin{minipage}{0.45\hsize}
  \begin{center}
   \includegraphics[bb=1 1 829 584, width=70mm]{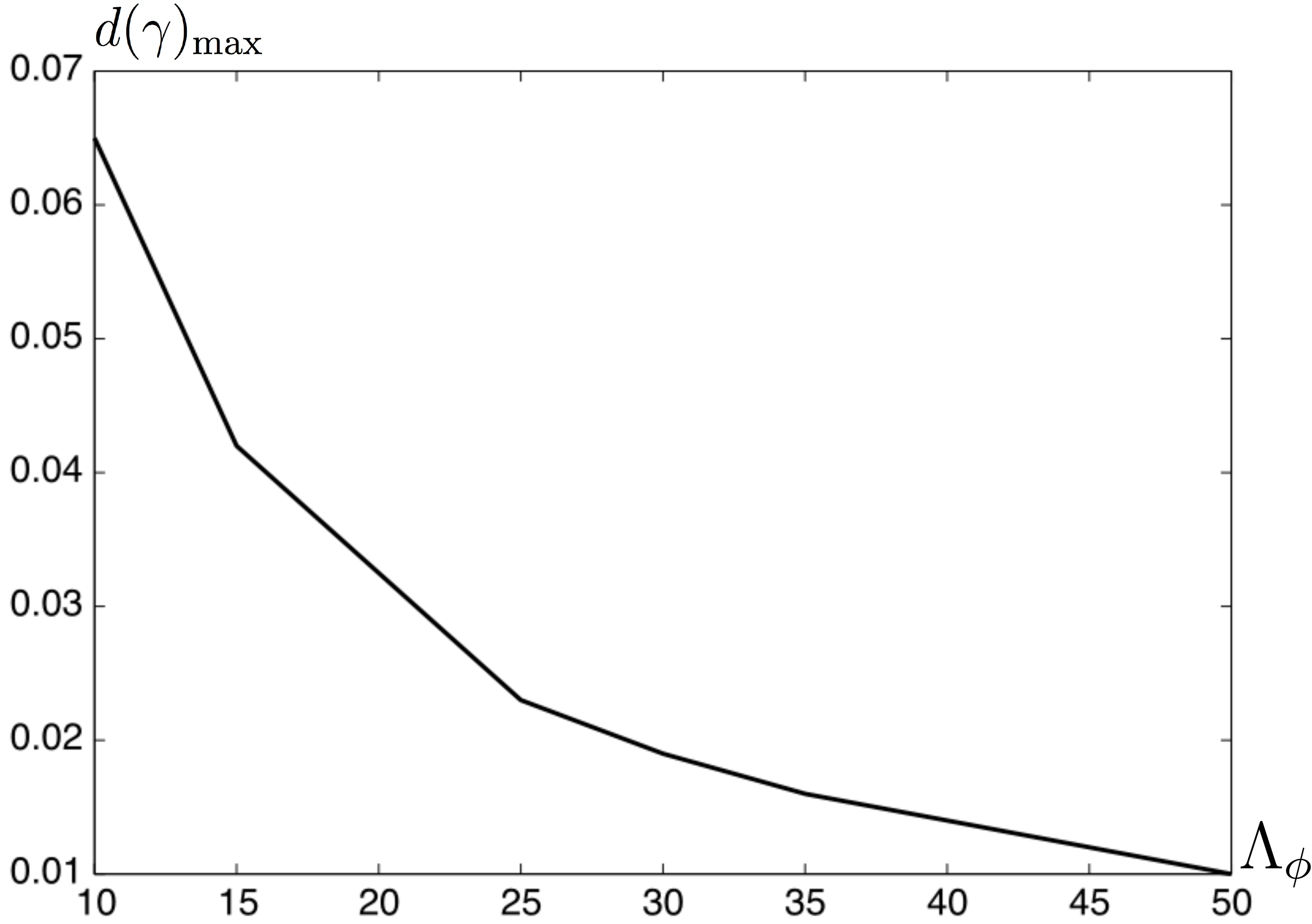}
   (a)
  \end{center}
 \end{minipage}
 \begin{minipage}{0.45\hsize}
  \begin{center}
   \includegraphics[bb= 0 1 820 586, width=70mm]{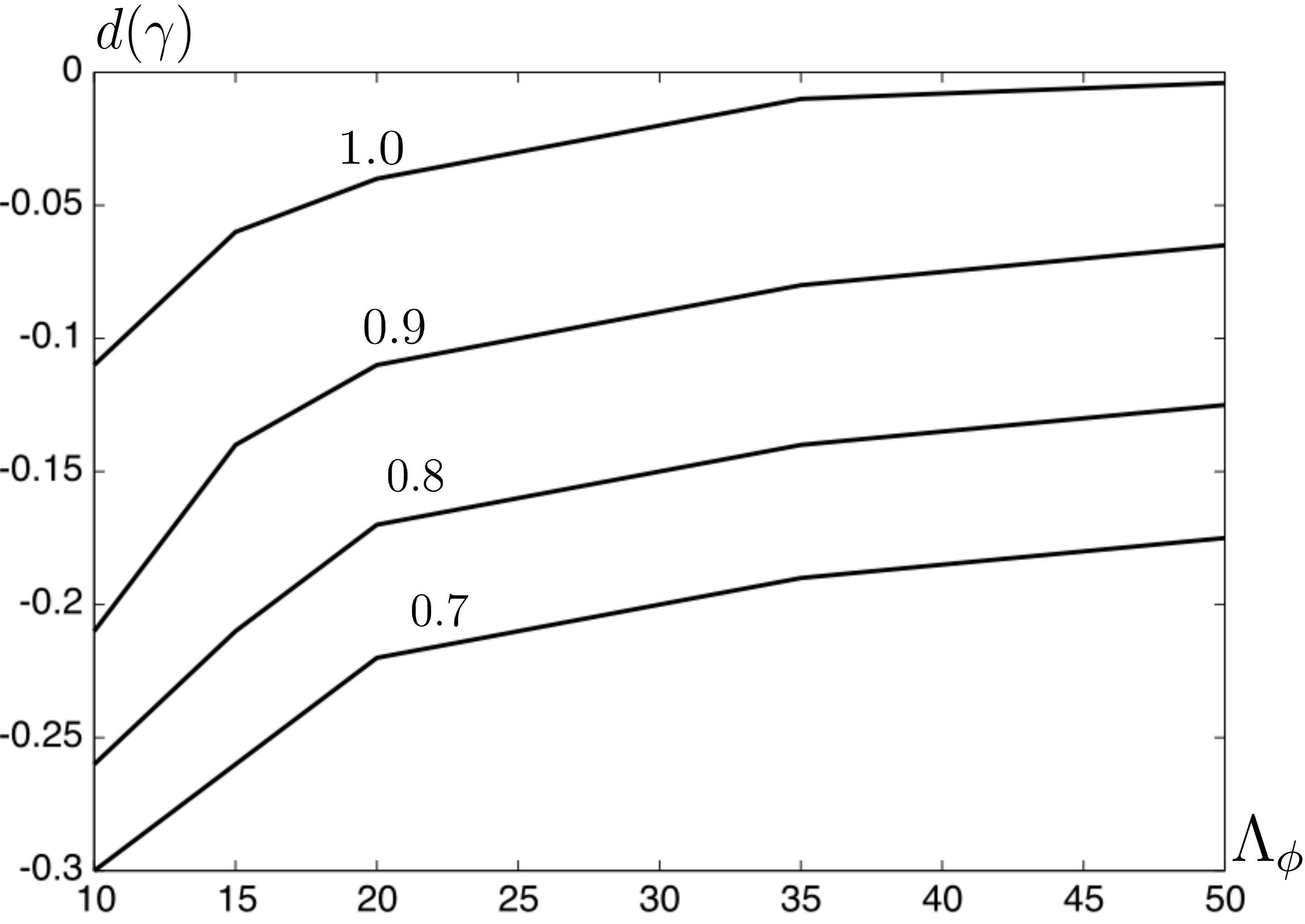}
   (c)
  \end{center}
  \end{minipage}

 \begin{minipage}{0.45\hsize}
\begin{center}
\includegraphics[bb=0 0 827 580, width=70mm]{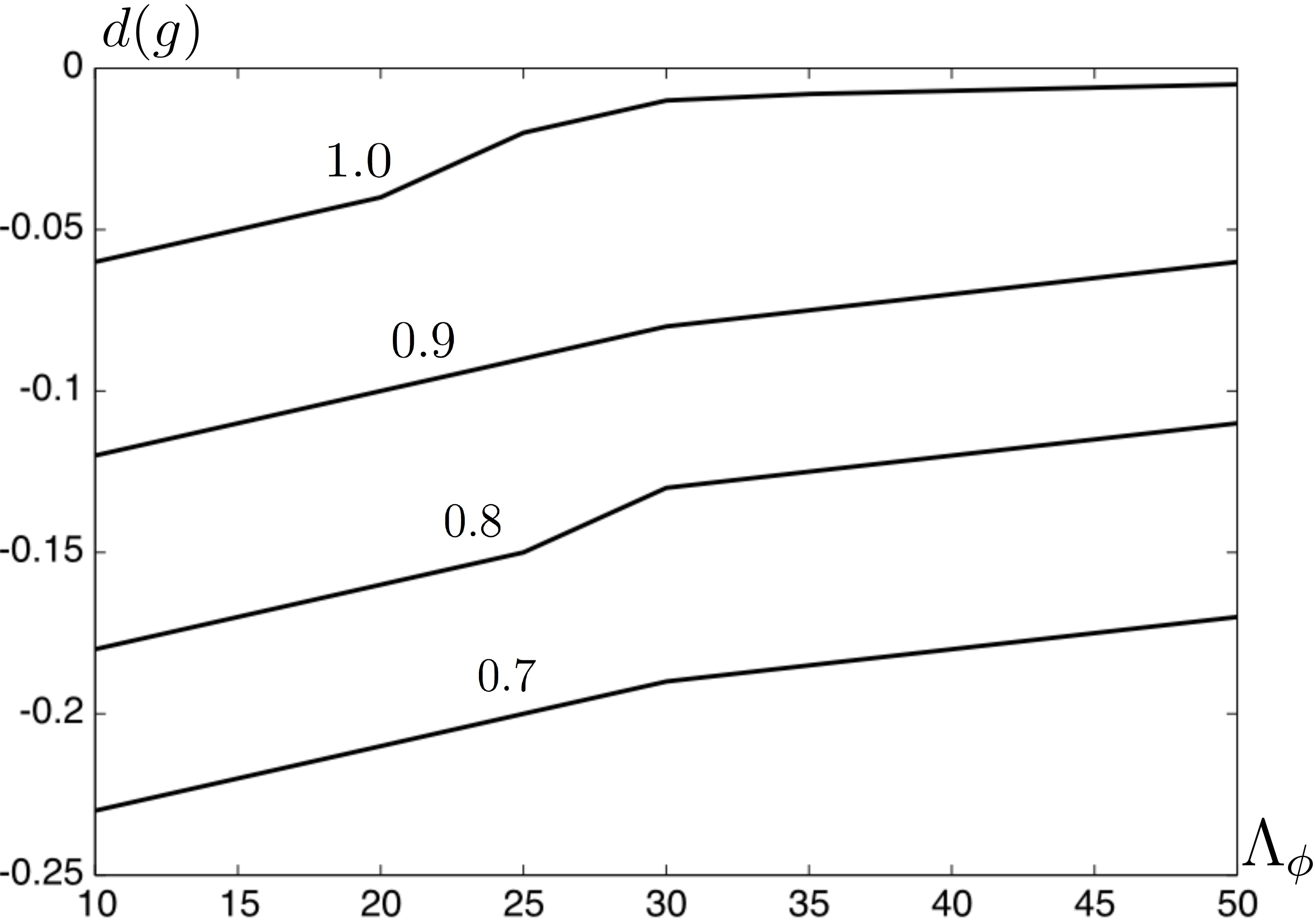}
(b)
 \end{center}
 \end{minipage}
 \begin{minipage}{0.45\hsize}
\begin{center}
\includegraphics[bb=8 0 861 633, width=70mm]{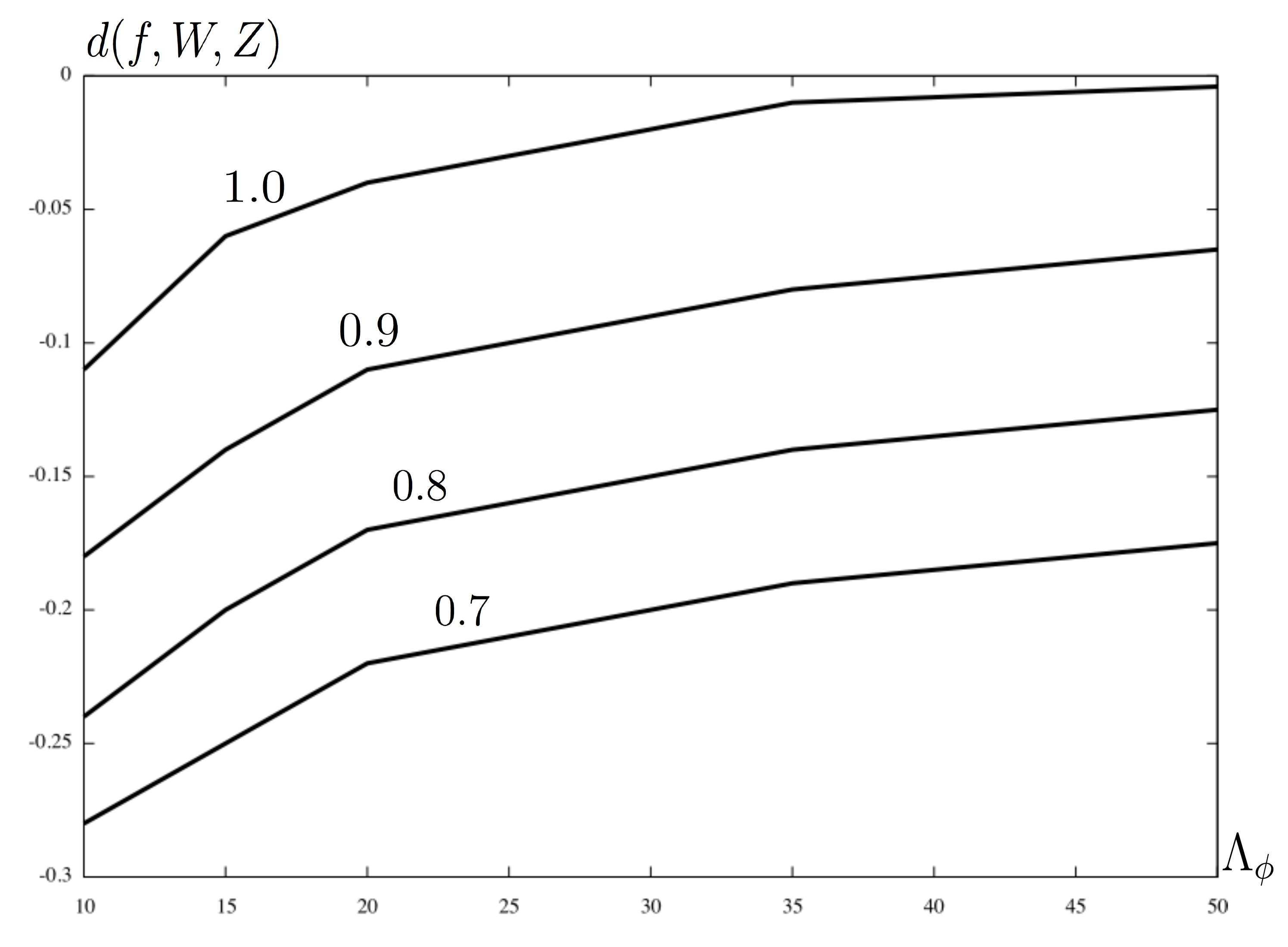}
(d)
\end{center}
 \end{minipage}
  \caption{The maximal deviations of $d(A)$, where d(A) is (a) $d(\gamma)>0$, (b) $d(g)$, (c) $d(\gamma)$ and (a) $d(f,W,Z)$  vs $\Lambda_\phi$ at $DR_{h_m}(\gamma\gamma)=0.7,0.8,0.9,1.0$.}
\label{fig:minmax}  
\end{figure}

When $\Lambda_\phi$ is bigger than 10 TeV, the deviation of the couplings because of the KK contributions of the loop and the Higgs-radion mixing becomes small.
In this section we consider a relation between the maximal  $|d(A)|$ and $\Lambda_{\phi}$.
For this purpose, we show the maximal value of $d(\gamma)$ in our model parameter scan 
as a function of $\Lambda_{\phi}$ in Fig~\ref{fig:minmax} (a) and the minimal value of $d(g)$,         $d(\gamma)$ and $d(f,W,Z)$  as a function of $\Lambda_{\phi}$ in Fig.~\ref{fig:minmax} (b),(c), and (d). 
In Fig.~\ref{fig:minmax} (b),(c) and (d), the minimum values of for $DR_{h_m} (\gamma\gamma)>0.7,0.8,0.9,1.0$ are shown, because by the time the ILC is built, $DR_{h_m} (\gamma\gamma)$ would be measured with a precision of $8\%(4\%)$.

Since the Higgs-radion mixing always suppresses $d(\gamma)$, the mixing parameter $\xi$ at $d(\gamma)_{max}$ is nearly zero.
The enhancement of $d(\gamma)$ comes from the KK contributions of quarks, leptons and the $W$ boson. Large enhancement appears when $F_{KK}$ is large because $d(\gamma)_{\rm max}$ is strongly suppressed with increasing $\Lambda_\phi$ in Fig.~\ref{fig:minmax} (a). The KK contributions are suppressed as 
$F_q^{KK} \propto 1/\Lambda_{\phi}$ and $F_W^{KK} \propto 1/\Lambda_{\phi}^2$.  
Even assuming the coupling measurement at the ILC500up (4\%), there is no observable model point for $\Lambda_\phi > 15$ TeV when $d(\gamma)>0$.
However, if we use the HL-LHC assuming CMS Scenario 2 together with the ILC500up, 
$d(\gamma)> 1.7$\% and we may be able to access up to $\Lambda_{\phi}=35$~TeV. 
For the ultimate sensitivity using the ILC1000up we can cover up to $\Lambda_{\phi}=50$ TeV.   
Of course, for higher $\Lambda_{\phi}$, the number of accessible model points becomes 
smaller and smaller.

The minimum value of $d(A)$ for $A=\gamma, g, (f,W,Z)$ depends on both the Higgs-radion mixing and the KK contributions.   
The Higgs-radion mixing universally suppresses the couplings between the 125 GeV scalar $h_{m}$ and the SM particles at the tree level.
The ratio between the detection ratios 
is useful to separate the Higgs-radion mixing effect from the KK contribution and will be discussed in Section V.  

First, we consider the $h_m  \rightarrow \gamma\gamma$ channel.
In Table.~\ref{Higgscouplingprecisions}, the error of $g(h\gamma\gamma)$ is 
$\sim 4 \%$ at the LHC 14 TeV.  At the ILC only, the error of $g(h\gamma\gamma)$ is limited by statistics.  When combined with $Br(h_m\rightarrow \gamma \gamma)$ 
 $/Br(h_m\rightarrow ZZ)$ measurements at the LHC, the sensitivity of the coupling could go down to  1.8\% at the ILC500up and less than 1\% at the ILC1000up. 
When $DR_{h_m}(\gamma\gamma)=0.9 $, which is 
within a $2 \sigma$  deviation from the SM at the HL-LHC, 
there are model points where deviation of $d(\gamma)$ exceeds the experimental sensitivity of 
the ILC500 for $\Lambda_{\phi} < 50$ TeV. 
When $DR_{h_m}(\gamma\gamma)$=1.0, the upper limit of the accessible scale becomes 20 TeV.
The model points where the expected deviation is comparable to the errors at LHC1000up exist even at 50 TeV.
The error of the coupling measurements is 4\%(2\%, 0.56\%) for $g(hgg)$ and 5\%(0.5\%, 0.21\%) for $g(hZZ)$ at the LHC 300 fb$^{-1}$ (ILC500, 1000up).  
 For  $DR_{h_m} (\gamma\gamma)\sim 0.9$,  there are model points with the deviation comparable to the experimental sensitivity exists up to  $\Lambda_\phi = 50$ TeV for $d(g)$
and $d(Z)$.

\section{Step toward the parameter determination} 
\begin{figure}
\includegraphics[bb=0 0 781 512, width=7cm]{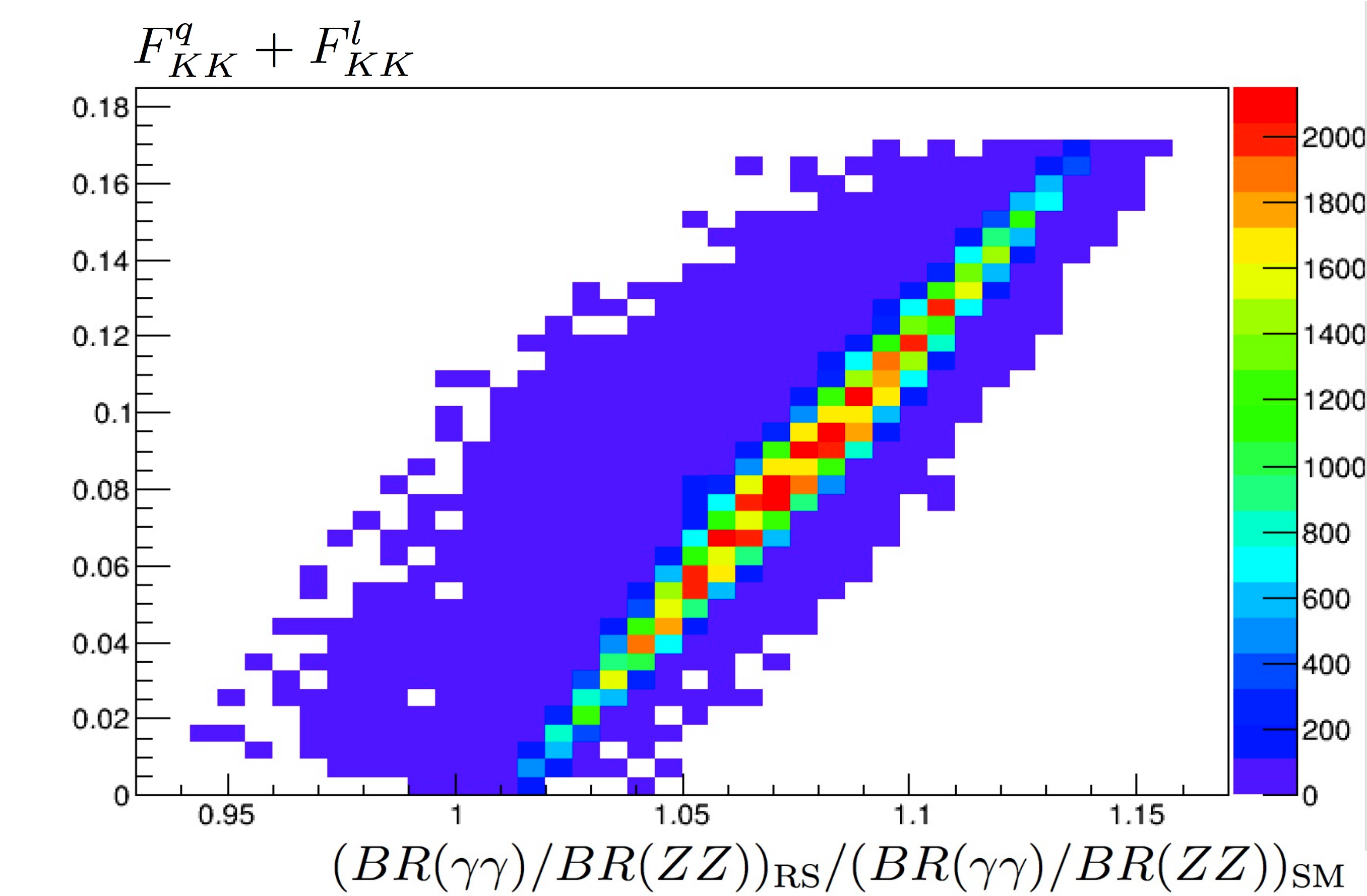}
\caption{$F^{KK}_l +F^{KK}_q$ vs the ratios of the branching ratio for $R(\gamma\gamma/ZZ)$ in all of our model points.  }\label{fkkvsratio}
\end{figure}

As we discussed already, the important parameters of our model 
affecting the Higgs boson branching ratios are
$F^{KK}_{q,l }$ and the Higgs-radion mixing controlled by $\xi$. At the LHC, the detection ratios $DR_{h_m }(A)$ can be measured directly. 
The ratio $R(\gamma\gamma/ZZ)\equiv DR_{h_m}(\gamma\gamma)/DR_{h_m} ( ZZ) $ is useful to determine $F_{KK}$ because both $\sigma( H\rightarrow gg)$ and 
the effect of the mixing in $DR_{h_m}(H\rightarrow \gamma \gamma, ZZ)$ cancel out in the ratio. 
Because of this, the ratio $R(\gamma\gamma/ZZ)$ is a sensitive function of $F^{KK}_q+ F^{KK}_l$.
This can be seen in  Fig.~\ref{fkkvsratio}, where our model points are 
clustered on a line. 
The model points which do not satisfy the linear relation correspond to 
small $m_r$ and large $|\xi|$, namely the points where the Higgs boson has a large mixing with a radion.  
This is an exceptional effect of the radion coupling to photons and gluons through anomaly in Eq. \ref{QCDanomaly}. 

\begin{figure}[htbp]
 \begin{minipage}{0.45\hsize}
  \begin{center}
   \includegraphics[bb=0 0 512 382, width=70mm]{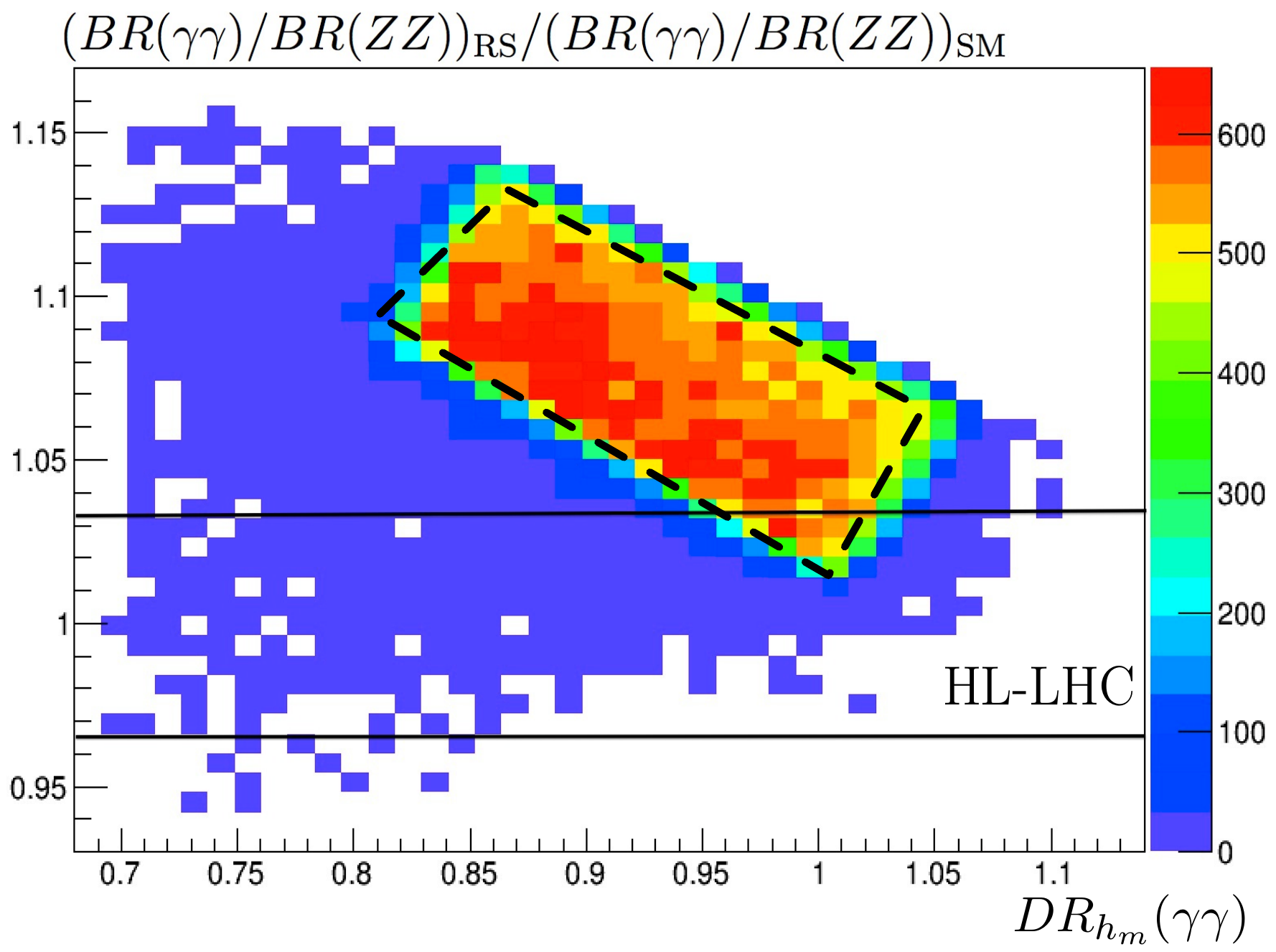}
   (a)
  \end{center}
 \end{minipage}
 \begin{minipage}{0.45\hsize}
  \begin{center}
   \includegraphics[bb=0 0 734 506, width=70mm]{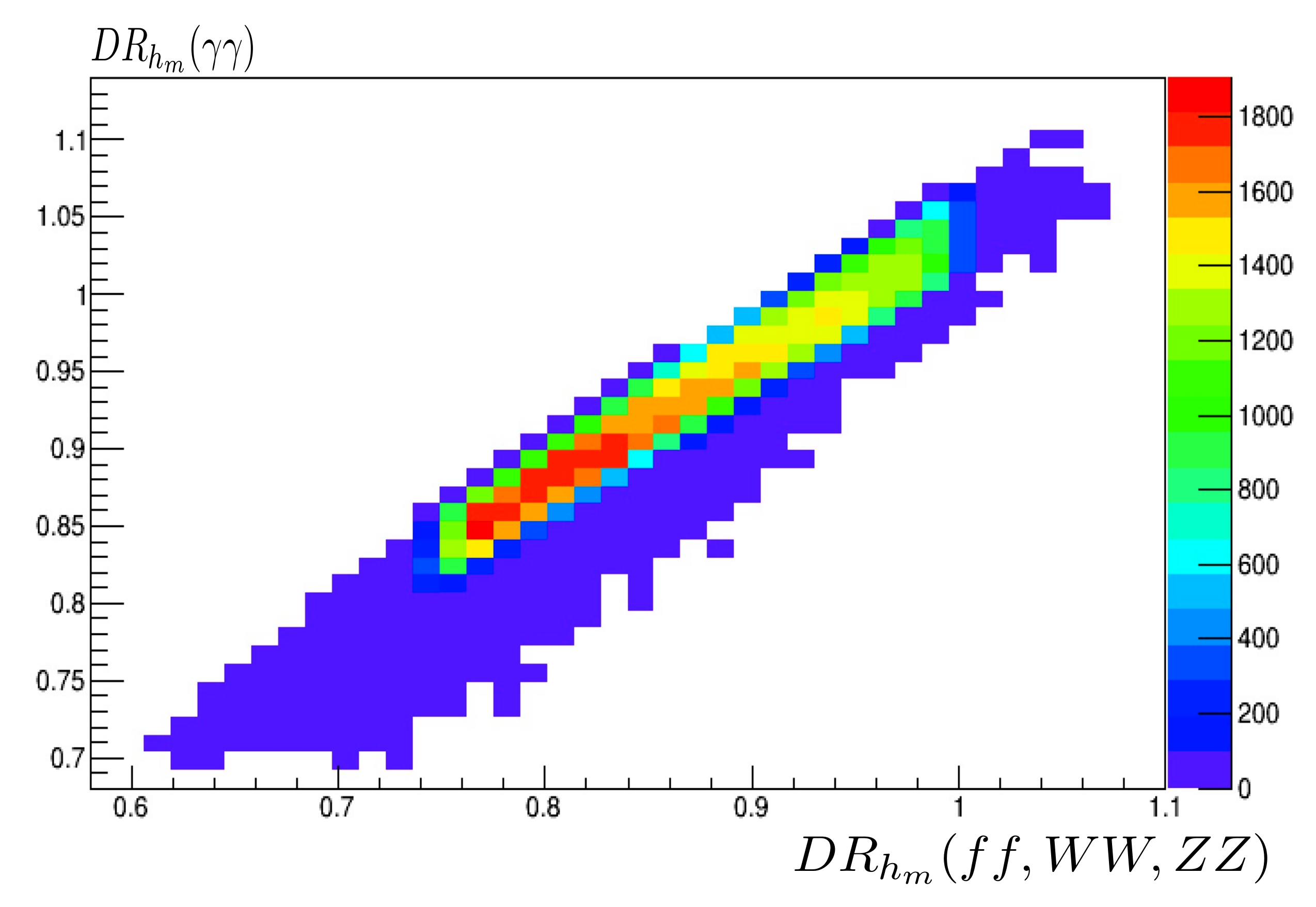}
   (b)
  \end{center}
 \end{minipage}
  \caption{(a) The distribution of the model points in $R(\gamma\gamma/ZZ)$  and $DR_{h_m}(\gamma\gamma)$ plane. The horizontal lines in the figure shows $1\sigma$ error centered at $R(\gamma\gamma/ZZ)=1$  at the HL-LHC; (b) The distribution of the model points between $DR_{h_m}(ff, WW ,ZZ)$ and $DR_{h_m}(\gamma\gamma)$. }
 \label{LHCquantity}
\end{figure}

\begin{figure}
\includegraphics[bb=9 0  815 561, width=15cm]{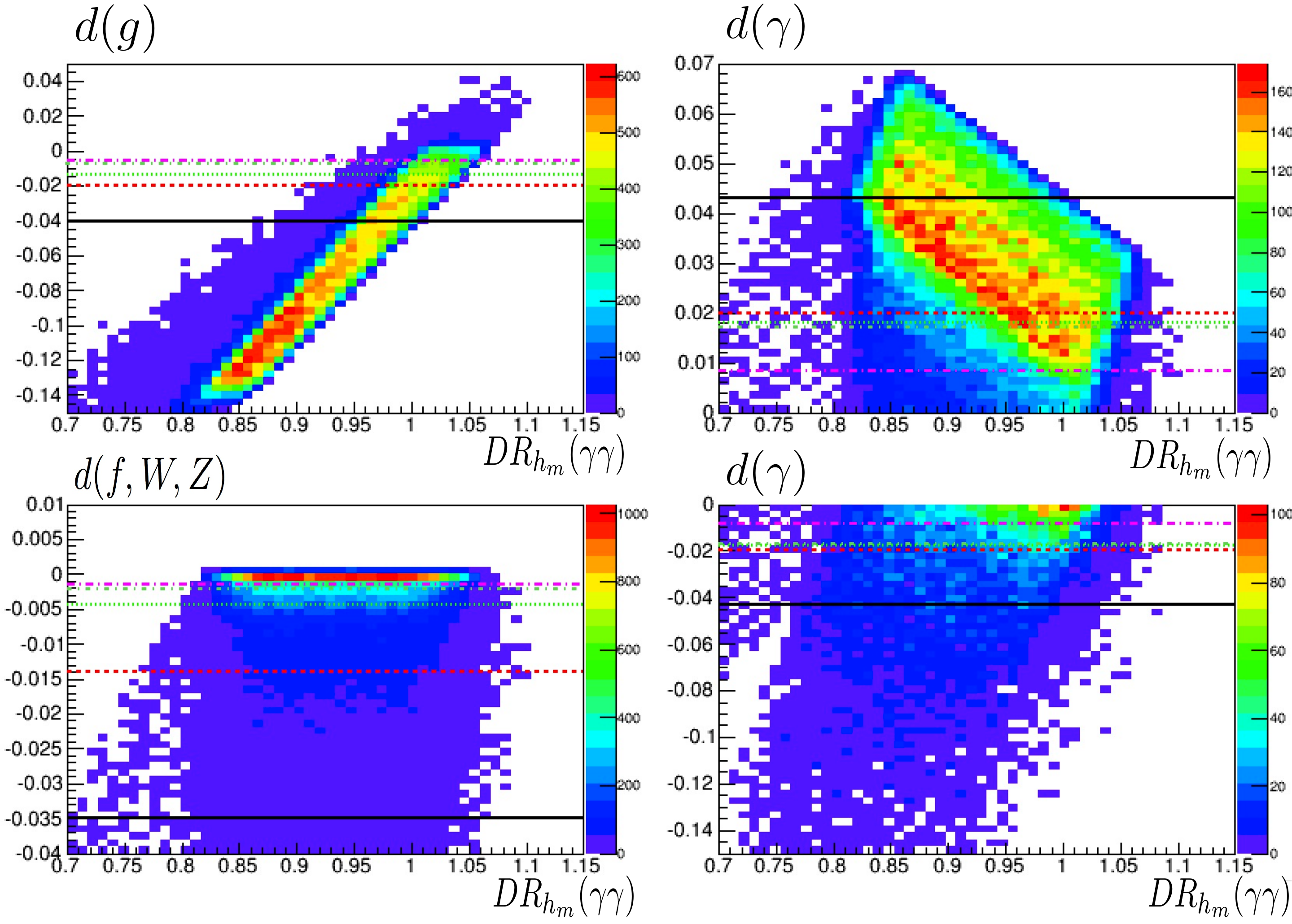}
\caption{Distributions of our model points in $DR_{h_m}(\gamma\gamma)$-$d(A)$ plane,
where $d(A)$ is (a) $d(g)$, (b) $d(f,W,Z)$, (c) $d(\gamma)$ with $d(\gamma)>0$ and (d)   $d(\gamma)$ with $d(\gamma)<0$. Expected $1\sigma$ sensitivity of the each coupling is also shown in the figure(see text).}
\label{fourplane} 
\end{figure}

We now show the correlation between $R(\gamma\gamma / ZZ)$ 
and $DR_{h_m}(\gamma\gamma)$ in Fig.~\ref{LHCquantity} (a).  
The Higgs-radion mixing is small for most of our model points, and they fall in the region indicated by the square box.
Small $DR_{h_m} (\gamma\gamma)$ is allowed when the Higgs-radion mixing is large.  
The horizontal lines in the plot 
correspond to the HL-LHC error as estimated by CMS~\cite{Peskin:2013xra}. 
Most of the model points of our scan predict deviations larger than the error. 
However, the fraction of model parameters with a more than $3 \sigma$ deviation 
from the SM prediction is small.
As we have already argued, some of the common systematics in $DR_{h_m}(\gamma\gamma)$ and $DR_{h_m}(ZZ)$ would be cancelled in the ratio.
Therefore the $R(\gamma\gamma/ZZ)$ measurement provides a reliable estimate of $F^{KK}_{q,l}$. 
In the Fig~\ref{LHCquantity} (b), we see the distribution of model points 
in the $DR_{h_m}(\gamma\gamma)$ -$DR_{h_m}( ff, WW, ZZ)$ plane. 
Most of the model points are distributed between $0.75<DR_{h_m}(ff,WW,ZZ)<1$, 
which is due to the correction to $\sigma(gg\rightarrow h_m)$. 
See Fig.~\ref{fourplane} for the coupling correction  $d(g)$ no Higgs-radion mixing at ($\xi\sim 0$),
where the model points range up to $d(g) \sim -0.15$.  

Currently,  the Higgs boson detection ratios are not strongly constrained. 
The error on $DR_{h_m }(\gamma \gamma)$ 
is about 30\% and the central values of ATLAS and CMS experiments are consistent with 
the SM expectation within $2 \sigma$. In the future the error may be reduced down to 4\% if the theoretical uncertainties can be controlled. 
In Fig.\ref{fourplane} we show our model points in the $DR_{hm}(\gamma\gamma)$
-$d(A)$ plane where  $d(A)$ is (a) $d(g)$, (b) $d(f,W, Z)$, (c) $d(\gamma)$ with $d(\gamma) > 0$ and (d) $d(\gamma)$ with  $d(\gamma)<0$.
In the plots, we also show the sensitivity  on the couplings estimated in \cite{Peskin:2013xra}  for 
LHC with 300 fb$^{-1}$(black solid line),  ILC250, 500, 500up and 1000up  from 
large $\vert \delta d(A) \vert $ to small which are listed in Table \ref{Higgscouplingprecisions}. 

At LHC 300 fb$^{-1}$, the number of accessible parameter points are not significant.  
At HL-LHC, 
the error will reduce by more than a factor of two. For the case of $d(g)$, most of the model points 
of our scan predict observable effects at the LHC, while deviations are rather small 
for the other couplings.

The improvement from the LHC to the ILC is significant in all channels.  For the case of 
$d(\gamma)$, 
there are  contributions of the KK $W$ bosons which is of the order of $1\%$ even if the contribution from $F^{KK}_{q, l }$'s are small.  
This will allow us to access most of the model parameter regions.
The deviation of the 
$d(f, W, Z)$ is expected to be small  except for the model points with small $m_{r_m}$.
However the expected sensitivity to $d(W)$ is also very good,
so that a significant parameter space can be covered by the ILC.

\begin{table}
\begin{tabular}{|c||c|c|c|c|c|c|c|c|c|}
\hline 
point & $m_{r_m}$& $\xi$ &$F_u^{KK}$& $F_l^{KK}$& $d(\gamma)$(\%)&  $d(Z)(\%)$  &$d(g)(\%)$  &$DR_{h_m}(\gamma\gamma)$&$DR_{h_m}(ZZ) $\cr 
\hline 
I        & 252    & 0.55    & 0.043 & 0.025 & 2.9 & $-0.18$ & $-6.0$&  0.94 & 0.89\cr
II         &  246    & 1.5    & 0.037   & 0.014 &  0.1 & $-2.1$ & $-5.0$& 0.94 & 0.90 \cr
III      &    187   & 1.2  & 0.025  & 0.013 & 2.3 &$-3.5$& $-2.4$& 0.95 & 0.97\cr 
\hline 
\end{tabular}
\caption{The model parameters of Points I, II and III. }
\label{modelpoints}
\end{table}

\section{Example coupling determination at model points}
In this section, we study the parameter determination of the RS model at the future colliders. 
For this purpose, we choose three model points which are allowed under the present 
experimental constraints. 
The model parameters of the RS model are the Higgs-radion mixing parameter $\xi$, 
the contributions of KK quarks $F_q^{KK}$ and KK leptons $F_l^{KK}$, the suppression factor of the radion couplings $\Lambda_\phi$ and the radion mass $m_{r_m}$.
Among these model parameters, $\xi$ and $F_q^{KK}$ affect the $h_m$ couplings strongly. Therefore, we mainly consider the constraints on  these two model parameters at each model point. 
We fix $\Lambda_\phi=10$ TeV.

The three model points we study are listed in Table \ref{modelpoints}. 
At Point I, $g(h\gamma\gamma)$ is enhanced from the SM prediction, and the  deviation is larger than the estimated error at the ILC1000up. 
Note that the effects of the KK loops enhance $g(h_m\gamma\gamma)$ and suppress $g(h_mgg)$ while the Higgs-radion mixing universally suppresses  all the $h_m$ couplings.
Therefore, a parameter point with $d(\gamma)>\delta d(\gamma)>0$ is realized only when $\xi$ is small. 
Deviation $d(f,W,Z) \sim 0$ at Point I, and $d(g)$ and $d(\gamma)$ receive contributions from the KK loops only.  

At Point I$\hspace{-.1em}$I, $d(\gamma)$ is nearly zero  due to cancelation between a reduction by the Higgs-radion mixing and an enhancement by the KK loops.
The deviation $d(\gamma)$ is too small to be measured  even at the ILC1000up+ HL-LHC. 
On the other hand,  $|d(f,W,Z)| \gg \delta d(f,W,Z)$ due to the Higgs-radion mixing. 
Coupling $g(hgg)$ is suppressed by both the Higgs-radion mixing and the KK loop effects.

At Point I$\hspace{-.1em}$I$\hspace{-.1em}$I, $DR_{h_m}(\gamma\gamma)$ is suppressed due to the large mixing. 
Since KK contributions enhance $d(\gamma)$,
the effect of the Higgs-radion mixing has to be large. This is achieved because $m_{r_m}$ is lighter than in Points I and I$\hspace{-.1em}$I. 
Compared with the others channels, both $|d(f,W,Z)|$ and $|d(g)|$ are significantly large at Point I$\hspace{-.1em}$I$\hspace{-.1em}$I.

\begin{figure}[htbp]
 \begin{minipage}{0.30\hsize}
  \begin{center}
   \includegraphics[bb=0 0 854 609, width=45mm]{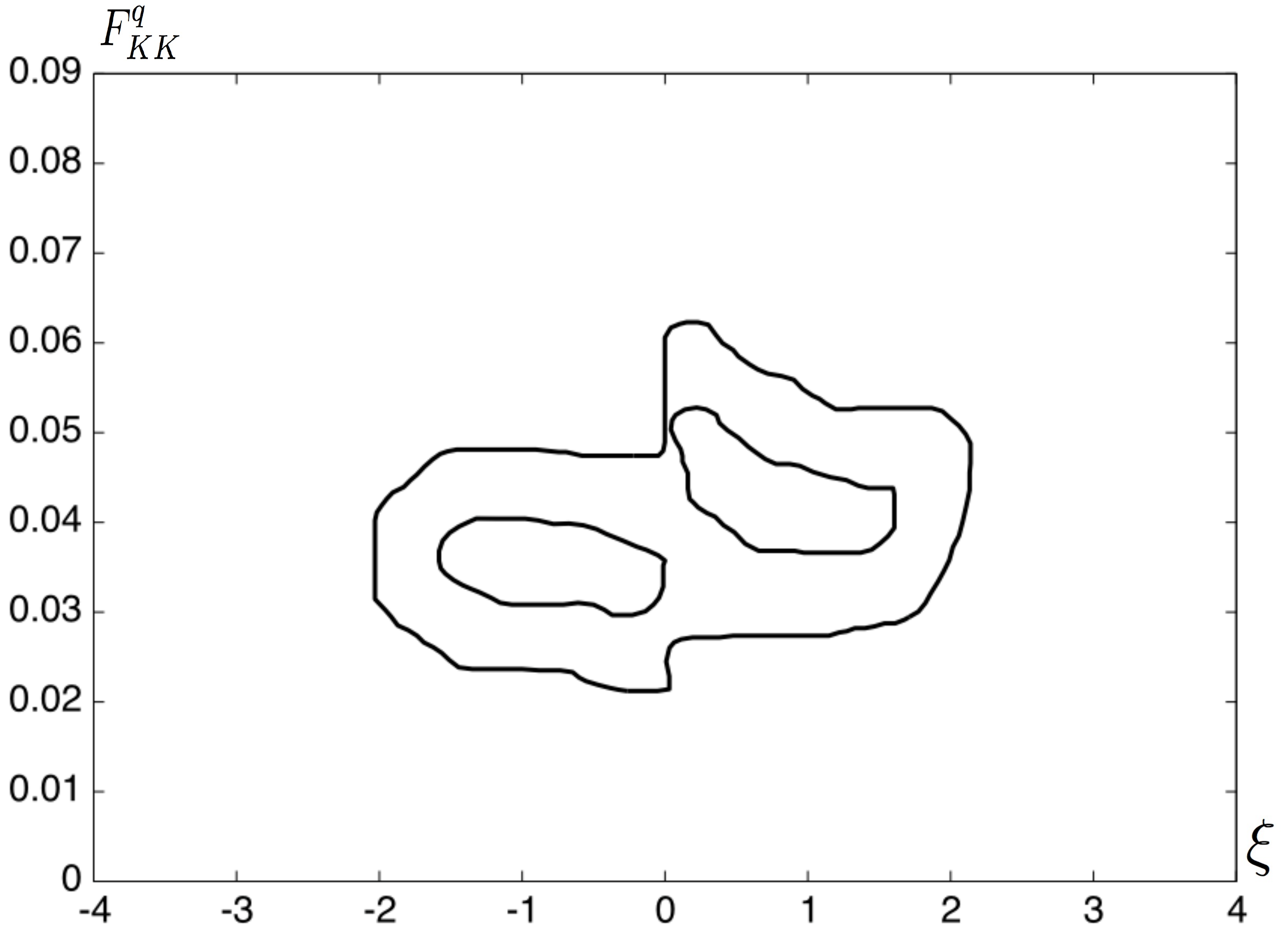}
   (a)
  \end{center}
 \end{minipage}
 \begin{minipage}{0.30\hsize}
  \begin{center}
   \includegraphics[bb=0 1 856 623, width=45mm]{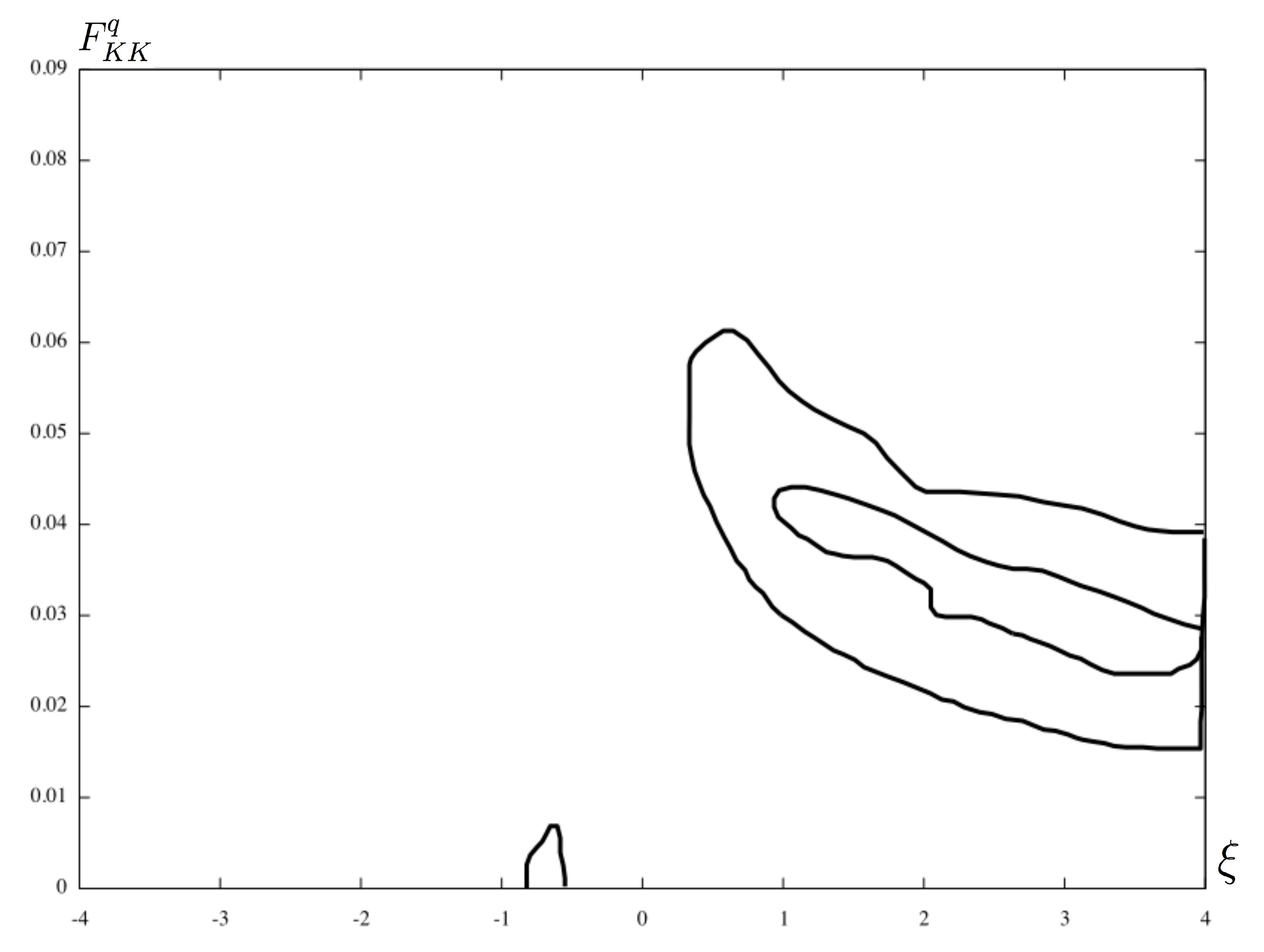}
   (b)
  \end{center}
 \end{minipage}
  \begin{minipage}{0.30\hsize}
  \begin{center}
   \includegraphics[bb=0 1 856 619, width=45mm]{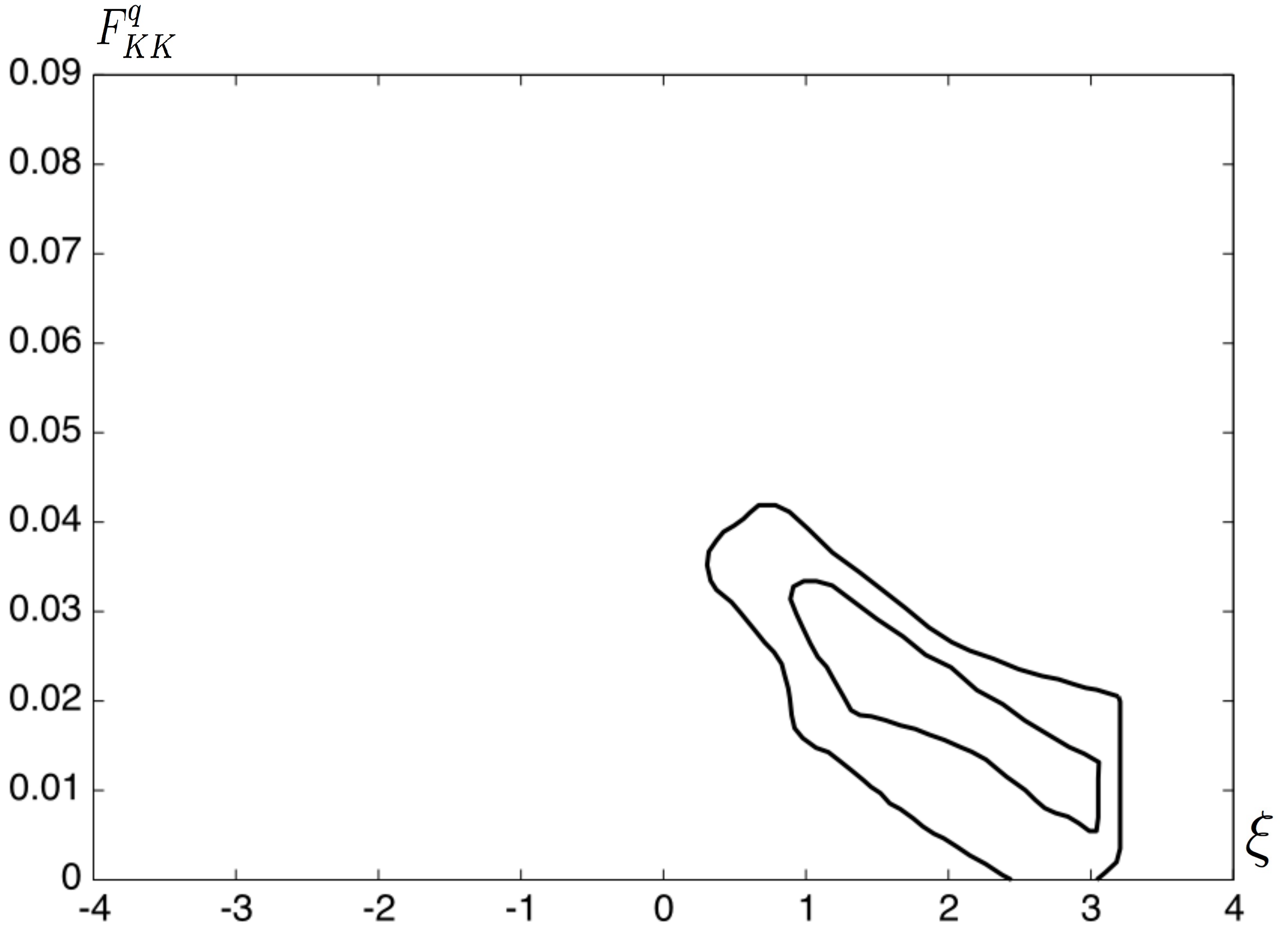}
   (c)
  \end{center}
 \end{minipage}
\caption{Contours of constant $\Delta\chi^2=1$ and $\Delta\chi^2=4$ in $F^q_{KK}$ and $\xi$ plane 
at (a) Point I,  (b) II and (c) III with $\Lambda_{\phi}=10$ TeV.
The $\Delta \chi^2$ is defined in Eq.~\ref{chidef} using the HL-LHC and ILC1000up statistics. }
\label{chi2contour}
\end{figure}

In Fig.~\ref{chi2contour} we plot the boundary of the model points in the  $(\xi, F_q^{KK})$ plane which 
satisfy $\Delta \chi^2<1$ and $\Delta \chi^2<4$, where $\Delta \chi^2$ is defined as follows,
\begin{equation}
\Delta\chi^2(p) \equiv \sum_i \frac{( Y_i(p)-Y_i(p_{\rm input } )^2}{(\Delta Y_i(SM))^2}. 
\label{chidef}
\end{equation}
Here $Y_i (p)=O_i(p)/O_i(SM)$, and $O_i(p)$ is a prediction of an independent observable  $O_i$ at a model point $p$, while 
$O_i(SM)$ is the SM prediction of $O_i$, 
and $\Delta Y^{SM}_i\equiv \Delta O_i(SM)/O_i(SM)$, where $\Delta O_i(SM)$  is the expected $1 \sigma$  
error of $O_i$ when $O_i= O_i(SM)$.  
The $O_i(p_{\rm{input}})$ is the value at our sample model points I, II or III.  
The functioin $\Delta \chi^2(p)$ is close to the expected 
 $\Delta \chi^2$ at point $p$ as far as $O_i(SM) \sim O_i(p_{\rm{input}})$, which is a good approximation for our model points. 
Here, the data and errors are  those of the ILC (Tables 5.4 and 5.5 of \cite{Asner:2013psa} ),  
and of CMS-HL-LHC Scenario 2  in \cite{Peskin:2013xra}, and a 3.6$\%$ error for the measurement of
$DR_{h_m}(\gamma\gamma) /DR_{h_m} (ZZ)$ as suggested in \cite{Peskin:2013xra}.

\begin{figure}[htbp]
 \begin{minipage}{0.7\hsize}
  \begin{center}
   \includegraphics[bb=0 0 660 460, width=70mm]{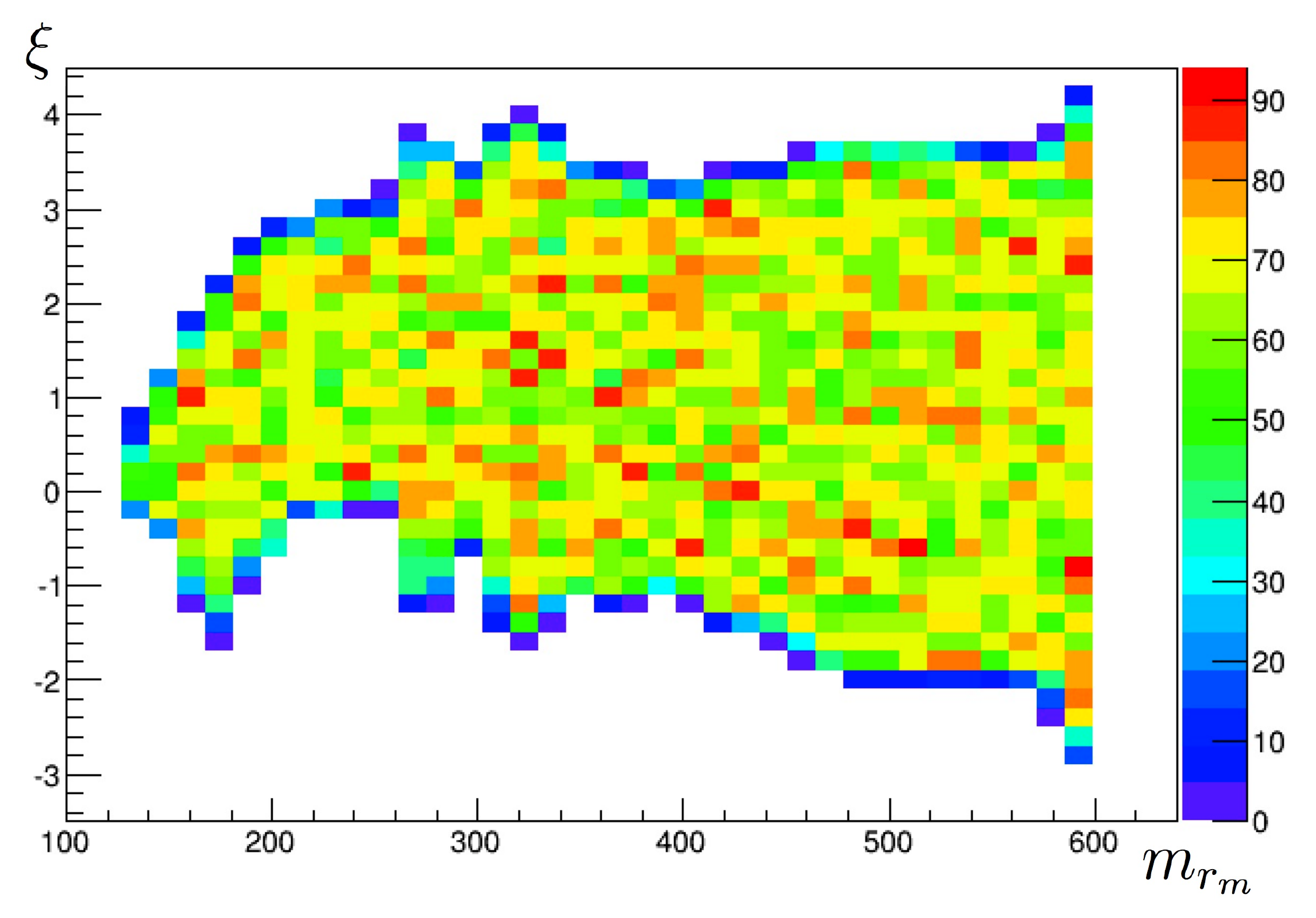}
  \end{center}
 \end{minipage}
\caption{Distributions of  our model points after applying the constraints from the
direct Higgs boson searches at the ATLAS and CMS experiments to the $r_m$ couplings.}
\label{xirm}
\end{figure}

The $\Delta \chi^2$ contour for Point I has two minima
corresponding to positive and negative $\xi$ solutions. 
Negative $\xi$ solutions do not appear at Point II and Point III because we 
include the constraints from direct Higgs boson searches at the LHC for $r_m$ couplings.  
In Fig.\ref{xirm} we show the distribution of model points after applying the direct Higgs 
search constraints in the $\xi$ and $m_{r_m}$ plane. 
In low $m_{r_m}$ region, $\xi$ is strongly restricted to reduce the couplings between the $r_{m}$ and SM particles.
The constraint is especially strong for $\xi<0$. 
 At Point II and Point III, we find the correlation between $\xi$ and $F_{KK}^q$. It arises due to the undetermined $m_r$ mass which would be determined elsewhere. For a given $\xi$, the best fit $r_m$ changes significantly in the plots.

One may also ask that wether it is possible to determine $\Lambda_{\phi}$ from the measurements. 
We found that the sensitivity is low because the effect of $\Lambda_{\phi}$ appears through $F^{KK}_{q,l}$  or the Higgs-radion mixing angle. 
When $\Lambda_{\phi}$ is increased, the maximal value of 
$\vert F_{KK} \vert $  reduces $\propto \Lambda_{\phi}^{-1}$. 
Therefore, $\Lambda_{\phi}$ will have an upper bound if  the measured $F^{KK}_{q,l}$ exceed their maximal values at the $\Lambda_\phi$. 
The maximal allowed value of $F^{KK}_q$ is 0.09 at $\Lambda_\phi=10$ TeV. 
However, the undetermined direction in the
$\xi$-$F^{KK}_{q,l}$ plane due to the unconstrained $m_{r_m}$ reduces sensitivity to the $\Lambda_{\phi}$ especially 
when $\xi$ is large.  
For Point I, the minimum $\Delta \chi^2=30$  at $\Lambda_{\phi}=50$~TeV, while acceptable parameter regions at $\Lambda_{\phi}=50$~TeV are still found for Point II and Point III.

\section{Discovery of the radion like state at the future colliders. }
 
\begin{figure}
 \begin{minipage}{0.45\hsize}
 \begin{center}
\includegraphics[bb=0 0 663 404, width=65mm]{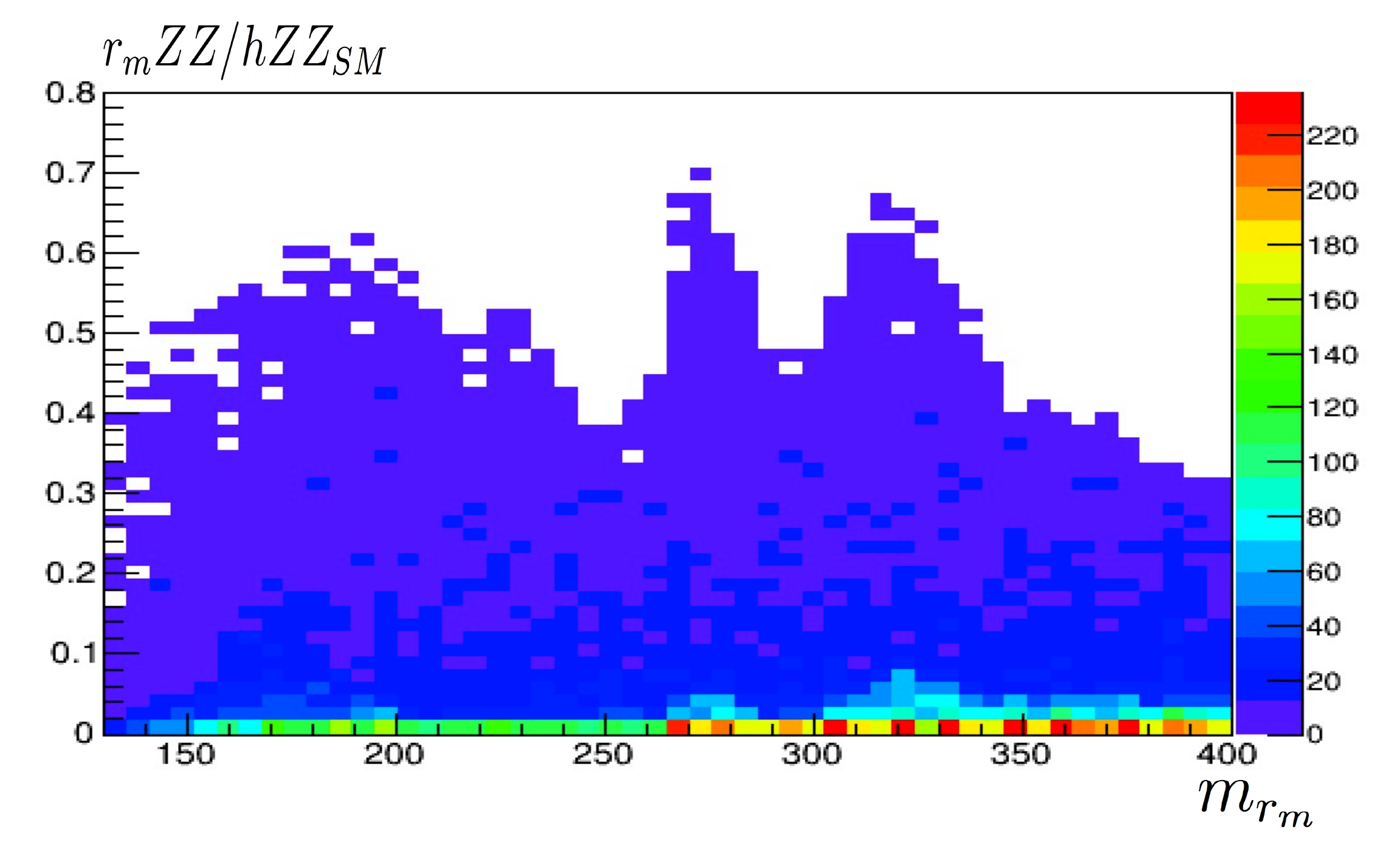}

(a)
\end{center}
\end{minipage}
\hskip 1cm
\begin{minipage}{0.45\hsize}
\begin{center}
\includegraphics[bb=0 0 678 504, width=66mm]{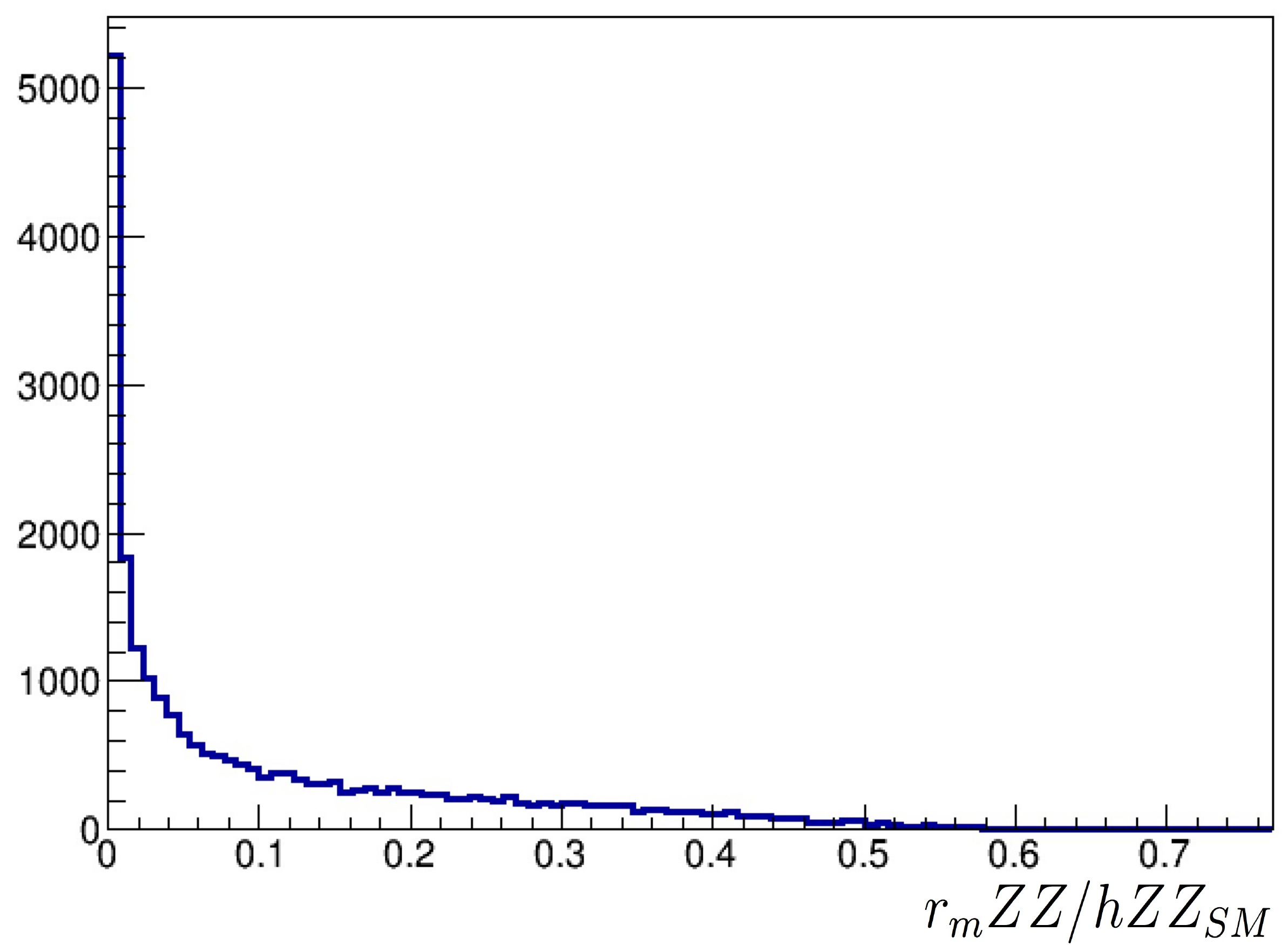}

(b)
 \end{center}
\end{minipage}
\caption{(a) The distribution of our model points after applying the LHC constraint in $m_{r_m}$ vs $d_{r_m}(Z) \equiv g(r_mZZ)/g(HZZ)_{SM}$ plane; (b) Number of model points vs $d_{r_m}(Z)$.}
\label{rzz}
  \end{figure}

In our scan, we restrict 130 GeV$<m_{r_m}<600$ GeV. Therefore, $r_m$ is 
kinematically accessible  at the proposed ILC, for which the current constraints at the LHC are relatively weak.
The radion-like state $r_m$ might be accessible at the ILC because the  production cross section could be  large enough if the $r_{m}$ mixes with the Higgs boson significantly. 
The distribution of model points in our scan in $d_{r_m}(Z) \equiv g(r_mZZ)/g(HZZ)_{SM}$ vs $m_{r_m}$ plane is shown in Fig.~\ref{rzz} (a) for $r_m<400$~GeV.
The maximal $d_{r_m}(Z)$ is limited by the current $h\rightarrow Z^0 Z^0$ search at the LHC, though 
the number of expected signal events is also  affected by  $d_{r_m}(g)$ through the $r_m$ production cross section.  For the scan at $\Lambda_\phi=10$~TeV, we found 24369 model points with $m_r<400$~GeV remain under the constraint, and $28\%$ of the model points satisfy $d_{r_m}(Z)>0.15$.  
The distribution of model points as a function of $d_{r_m}(Z)$ is shown in Fig.~\ref{rzz} (b).

The branching ratio of $r_m$ changes drastically in the kinematically accessible range of $m_{r}$. 
The dominant branching ratio of $r_m$ is the same as that of the Higgs boson with the same mass, because the coupling arises from the Higgs-radion mixing when $\Lambda_{\phi}$ is an large. Above $r_m>200$~GeV, the  decay  $r_m \rightarrow Z^0Z^0$ has a significant branching ratio and it is an important  discovery channel at the ILC. For $r_m> 135$ GeV, the decay $r_m\rightarrow W^+W^-$ has the largest branching ratio.  If the $r_m$ mass is smaller than this value, the decay $r_m\rightarrow b\bar{b}$ is dominant. 

The sensitivity of the ILC to $r_m$ may be estimated from the existing studies on Higgs boson 
searches at the ILC.  
The searches using the decay mode $H \rightarrow W^+W^-$ and  $Z^0Z^0$ can be found in Ref~\cite{Meyer:2003ts} . The number of signal and background events for $e^+e^-\rightarrow ZH\rightarrow 2l 4q$ at the ILC with $\sqrt{s}=500$~GeV and $\int dt {\cal L}=500$ fb$^{-1}$ are listed in this reference. 
After all kinematical cuts, the numbers of signal events for $m_H =200, 240, 280, 320$~GeV are $585, 463, 333, 216$ while the number of background events is  $275$. 
From those numbers it is possible to set a $2 \sigma$ exclusion limit without systematical uncertainty as $d_{r_m} (Z)<0.24,0.28, 0.33,0.40$  for $m_h  =$ $200,240,280, 320$~GeV respectively.   In Fig.7 of Ref.~\cite{Meyer:2003ts}, the signal distributions are seen on top of the background and the combinatoric events of the signal. Taking events with the reconstructed di-boson mass $m_{VV}$ within $m_h \pm 10$~GeV, we obtain an alternative $2 \sigma$  exclusion limit $d_{r_m}(Z)<0.20$, 0.26, 0.32, 041  for $m_H  =$ $200,240,280, 320$~GeV respectively.

The Higgs boson  with a mass between 135 GeV and 180~GeV decays dominantly to the $W^+W^-$ final state. A study has been done in Ref.~\cite{GarciaAbia:2000bk} for the $llW^+W^-$ channel and for $qq W^+W^-$ followed by the hadronic decay of $W^{\pm }$ assuming $\int dt L = 500$ fb$^{-1}$  at $\sqrt{s}=350$~GeV.  The signal and background distributions are given in Fig. 2 of \cite{GarciaAbia:2000bk} for the SM Higgs boson with a mass between 150 GeV and 180 GeV. In the peak region, we find $\sim 510(1205)$ signal events over $\sim 100(519)$ background events for $m_H=150(180)$ GeV in the $llW^+W^-$ ($qq W^+W^-$) channel respectively.  This means a $2\sigma$ exclusion of  $d_{r_m}(Z)> 21(20)$\% is possible. 
The estimation here may be conservative because we do not take into account the other decay 
channels.

For $m_H<135$ GeV, the dominant decay mode is $H\rightarrow b\bar{b}$. In Ref.~\cite{Ono:2012mm}, the numbers of the signal and the background events are given for various 
 processes at $\sqrt{s}=250$ GeV and $\int dt  L=250$fb$^{-1}$  using a polarized electron beam as shown in Table~\ref{hbb}.
 
\begin{table}
 \begin{tabular}{|c||c|c|c|}
\hline 
                       &signal ($m_H=120$~GeV)& background   \cr
\hline
 $\nu\nu H $      &  4753  &   3593    \cr               
$ qqH $               & 13726  &  166807          \cr
 $ll H (e)$            &  1184   &   1607 \cr
  $llH ( \mu)$      &   1365  &      638    \cr
 \hline 
 \end{tabular}
 \caption{Number of signal and background events in the $H \to bb$ channel  as given in Ref.~\cite{Ono:2012mm}.}
 \label{hbb}  
 \end{table}
 
 The expected $2\sigma$ exclusion from the $\nu \nu H$  channel is $ d_{r_m}(Z) >0.16$ for $m_H=120$ GeV without taking 
 systematic errors into account.  For the other higgs boson mass, the number of  signal events
 should be scaled by the signal cross section $\sigma_{H \rightarrow bb}$ (= ($\sigma(e^+e^- \rightarrow ZH)$) 
 $\times (Br(H \rightarrow b\bar{b}))$ and it reduces rapidly with the Higgs mass. 
The signal cross section  is 350 (275, 120) fb for  $m_h=110(120,135) $~GeV
 and the sensitivity to the $d_{r_m}(Z)$ should be scaled accordingly, namely  $d_{r_m}(Z)>0.24$ at $m_H=135$ GeV. 

As we have seen in Fig.~\ref{rzz} (b), the parameters with large Higgs-radion mixing are rather limited already 
and the discovery of $r_m$ is not guaranteed.
If $r_m$ is discovered, it will improve $F_{KK}$ determination from the Higgs coupling measurement by resolving the correlation with $\xi$, which is crucial to estimate the scale of $\Lambda_\phi$.

\section{conclusion }
In this paper, we have investigated the phenomenology of the Higgs-radion sector of the RS model in the light of the proposed ILC and HL-LHC. 
In this model, the lightest KK mass has to be above 10 TeV due to the kaon-mixing constraint.
Therefore, we can not discover the KK particles directly,
and the deviations of the Higgs couplings from the SM predictions may be small.
On the other hand, we can measure the Higgs couplings precisely at the ILC.
It is important to note
that $g(H\gamma\gamma)$ may be measured precisely when we combine the results of the ILC with that of the HL-LHC.

In this model, contributions of the new physics to the Higgs couplings are from Higgs-radion mixing and KK loops.
The Higgs-radion mixing commonly suppresses the Higgs couplings in all channels.
In the $h\gamma\gamma$ channel, there are contributions from the KK $W$ bosons, KK quarks and KK leptons in addition to the Higgs-radion mixing contribution. 
The KK contributions enhance $g(H\gamma\gamma)$, 
and the branching ratio can be above the SM prediction when the Higgs-radion mixing controlled by the mixing parameter $\xi$ is small.  
On the other hand, the KK contribution suppresses $g(hgg)$.  
The other couplings $g(hff), g(hWW)$ and $g(hZZ)$ receive suppression from the Higgs-radion mixing.

The RS model is an example in which new physics effect is expected in the Higgs sector.
The deviations of the 125 GeV higgs couplings to the SM particles $d(g), d(f, W, Z)$, $d(\gamma)$
may be large enough to be measured at the future colliders.
For some model parameters, the discovery of a radion-like mass eigenstate will significantly improve the sensitivity to those parameters.
The proposed ILC's will confront the theory at least up to a mass scale of the  $O(10)$ TeV, which is far beyond the expected sensitivity of 14 TeV.

For $\Lambda_\phi=10$ TeV, most of the model points of our scan predict deviation of Higgs boson couplings which is large enough to be detected at the proposed ILC and the HL-LHC.
When $\Lambda_\phi$ is increased above this scale, the coupling deviation reduces rather quickly. However, there are model points accessible at the future ILC even if $\Lambda_\phi=50$ TeV.

We also study parameter reconstruction at selected model points. The KK contribution and the mixing angle are precisely measured while the fundamental parameters such as $\Lambda_\phi$ and $\xi$ are difficult to determine. 
However if the radion state $r_m$ is also discovered, those parameters would be measured more precisely.

\section*{Acknowledgments}
This work is supported by Grant-in-Aid for Scientific research from the Ministry of
Education, Science, Sports, and Culture (MEXT), Japan (No.23104006 for M.M.N.), and also by World Premier International Research Center Initiative (WPI Initiative), MEXT, Japan.

\renewcommand{\refname}{References}

\end{document}